\documentclass[a4paper,11pt,headings=big,DIV=12]{scrartcl}
\pdfoutput=1
\usepackage{amsmath,amssymb,mathtools}
\usepackage[colorlinks,linkcolor=mygray,urlcolor=blue,citecolor=blue]{hyperref}
\usepackage{tcolorbox} 
 
\definecolor{mygray}{gray}{0.2}
 \addtokomafont{disposition}{\rmfamily\boldmath}
  \usepackage{bbold}

\definecolor{mypink1}{rgb}{0.9, 0.2, 0.6}
\usepackage{graphicx}
\addtokomafont{disposition}{\rmfamily\boldmath}
\usepackage[utf8]{inputenc}
\usepackage{enumerate} 
\usepackage{slashed}
\usepackage{cite} 
\usepackage{comment}
\usepackage{multirow}
\usepackage{acronym}
 \usepackage{bbold}
 \usepackage{environ}
\NewEnviron{thincases}{\scalebox{0.5}[1]{$\displaystyle
\left\{\scalebox{2}[1]{\setlength{\arraycolsep}{6pt}% <- I did not look up the "correct" value
$\displaystyle\begin{array}{ll}
\BODY
\end{array}$}\right.$}}%}
\NewEnviron{thinpmatrix}{\scalebox{0.5}[1]{$\displaystyle
\left(\scalebox{2}[1]{$\displaystyle
\begin{matrix}
\BODY
\end{matrix}$}
\right)$}}
\allowdisplaybreaks
\DeclareOldFontCommand{\rm}{\normalfont\rmfamily}{\mathrm}
\DeclareOldFontCommand{\sf}{\normalfont\sffamily}{\mathsf}
\DeclareOldFontCommand{\tt}{\normalfont\ttfamily}{\mathtt}
\DeclareOldFontCommand{\bf}{\normalfont\bfseries}{\mathbf}
\DeclareOldFontCommand{\it}{\normalfont\itshape}{\mathit}
\DeclareOldFontCommand{\sl}{\normalfont\slshape}{\@nomath\sl}
\DeclareOldFontCommand{\sc}{\normalfont\scshape}{\@nomath\sc}

\numberwithin{equation}{section}

\newcommand{\beS}{{b_{\scal}}}
\newcommand{\beF}{{b_{\ferm}}}

\newcommand{\Ima}{\textrm{Im}}

\newcommand{\Dechi}{ \De_{\bar qq}}

\newcommand{\vev}[1]{\langle #1 \rangle}

\newcommand{\matel}[3]{\langle #1|#2|#3\rangle}

\newcommand{\al}{\alpha}
\newcommand{\be}{\beta}
\newcommand{\ga}{\gamma}

\newcommand{\de}{\delta}
\newcommand{\De}{\Delta}
\newcommand{\la}{\lambda}

\newcommand{\Sig}{ \Sigma}
\newcommand{\sig}{ \sigma}

\newcommand{\GaC}{\Theta}

\newcommand{\EQ}{Eq.~}

\newcommand{\FIG}{Fig.~}

\newcommand{\SEC}{section~}
\newcommand{\SECs}{sections~}
\newcommand{\APP}{appendix~}

\newcommand{\cw}{\omega}
\newcommand{\Lag}{{\cal L}}
\newcommand{\LagImp}{{\cal L}^R}
\newcommand{\TImp}{T^R}

\newcommand{\xid}{\xi_d}

\newcommand{\xif}{\xi_4}

\newcommand{\dphi}{d_\varphi}
\newcommand{\gzero}{\Theta} 
\newcommand{\gone}{A}
\newcommand{\gtwo}{\cal D}
\newcommand{\gJ}{J}

\newcommand{\Lageff}{\Lag_{\text{eff}} }

\newcommand{\Lagkind}{\Lag_{\text{kin,d}}}

\newcommand{\XX}{\chi}

\newcommand{\potd}{V_d}

\newcommand{\Tr}{\mathrm{Tr}}

\newcommand{\ORD}{{\cal O}}

\newcommand{\gast}{\ga_*}

\newcommand{\Gtwo}{\gtwo}

\newcommand{\dil}{D}

\newcommand{\scal}{\varphi}
\newcommand{\ferm}{\psi}
\newcommand{\scalg}{\varphi}

\newcommand{\dbar}{{d-1}}
\newcommand{\TEMT}{ \Tud{\rho}{\rho} }
\newcommand{\Tud}[2]{T^{#1}_{\phantom{#1} #2}}

\newcommand{\mink}{\eta}

\newcommand{\pert}{\la}

\newcommand{\chiPT}{d$\chi$PT }

\definecolor{violet}{rgb}{0.94, 0.2, 0.8}%lightcoral
\definecolor{lightblue}{rgb}{0.39, 0.58, 0.93} %cornflowerblue
\definecolor{lightgreen}{rgb}{0.1, 0.73, 0.33}

\newcommand{\RZ}{}

\newcommand{\Wcov}{\De}

\DeclareOldFontCommand{\tt}{\normalfont\ttfamily}{\mathtt}

\begin{document}
\begin{flushright}
CERN-TH-2023-101 \\
\today
\end{flushright}

\begin{center}
{\Large\bfseries \boldmath Dilatons Improve (Non)-Goldstones}\\[0.8 cm]
{\Large%
 Roman Zwicky,
\\[0.7 cm]
\small
Higgs Centre for Theoretical Physics, School of Physics and
Astronomy,  \\ The University of Edinburgh,  
Peter Guthrie Tait Road, Edinburgh EH9 3FD, Scotland, UK \\[0.1cm]
Theoretical Physics Department, CERN,  \\
Esplanade des Particules 1,  Geneva CH-1211, Switzerland
} \\[0.7 cm]
\small
E-Mail:
\texttt{\href{mailto:roman.zwicky@ed.ac.uk}{roman.zwicky@ed.ac.uk}}.
\end{center}
\thispagestyle{empty}
\begin{abstract}
{
Shift symmetry forbids conformal coupling of Goldstone bosons from internal symmetries, but not for spontaneously broken conformal symmetry.
 Its Goldstone boson, the dilaton $D$, admits and indeed requires, an improvement term 
$ {\cal L}_R \propto R e^{-2D/F_D}$
as it realises the Goldstone matrix element in the effective theory.
The improvement, combined with Weyl-gauging, enables conformal  coupling 
to Goldstone bosons and other particles  of  arbitrary Weyl-weight.
While improvement does not affect scattering amplitudes in flat space, 
it impacts gravitational form factors decisively, giving rise  to the dilaton pole in the spin-zero channel. 
We compute leading-order scalar, fermion, pion, and dilaton form factors, confirming low-energy constraints.  
The dilaton decoupling limit further implies that the operator driving spontaneous chiral symmetry breaking has scaling dimension $\Delta= d-2$.
 }  
\end{abstract}

\setcounter{tocdepth}{3}
\setcounter{page}{1}
\pagestyle{plain}
\setcounter{page}{1}
\mdseries

\setcounter{page}{1}

\setcounter{footnote}{0}

\tableofcontents

\section{Introduction}

The free massless scalar field, minimally coupled  to  gravity, is not conformal unless the spacetime dimension is $d=2$. 
This leads to problematic  ultraviolet (UV) behaviour in $ d \neq 2$ quantum field theories   
\cite{Callan:1970ze,DeWitt:1975ys}, and is \RZ{found to be}  in tension with the weak equivalence principle
   \cite{Chernikov:1968zm}. 
These issues are resolved when the scalar field is non-minimally coupled  to the Ricci scalar \cite{Chernikov:1968zm,Callan:1970ze,DeWitt:1975ys}, restoring conformality.  
However, the shift symmetry  prohibits the improvement of
 Goldstone bosons due to a broken internal global symmetry   
  \cite{Voloshin:1982eb,Leutwyler:1989tn,Donoghue:1990xh,Donoghue:1991qv}.

 The dilaton $D$, understood 
as the Goldstone boson  due to spontaneous symmetry breaking of conformal symmetry \cite{Isham:1970gz,Coleman:1985rnk,Low:2001bw}, changes matters.  The essential point is that the dilaton itself can be improved 
and automatically improves the  remaining Goldstones as it realises the conformal symmetry (non-linearly). 
  In fact,  dilaton improvement 
is indispensable, as it realises the fundamental Goldstone matrix element 
$\matel{0}{T_{\mu\nu}}{\dil(q)} =  -{F_{\dil} }/(d-1) q_\mu q_\nu$ in the effective theory. 
We provide a systematic discussion of how the dilaton improvement, affects pions 
and ordinary particles. The latter involves Weyl-gauging since the conformal weights depend on the underlying theory and are not  necessarily the free-field values.

In the literature dilaton improvement is  sometimes considered 
\cite{Migdal:1982jp,Schwimmer:2010za,Komargodski:2011vj,Bobev:2013vta,Golterman:2016lsd,Zwicky:2023krx}  and sometimes not \cite{Coleman:1985rnk,Yamawaki:2015tmu,Appelquist:2022mjb} as 
it does  not affect scattering amplitudes in flat space.  However, it does
 impact gravitational form factors which we study at leading order (LO) for the systems discussed above.
In the pion case a low-energy theorem constrains   the operator breaking chiral 
symmetry to be of scaling dimension $d-2$, consistent with alternative methods 
\cite{Zwicky:2023bzk,Zwicky:2023krx}. 
Below, we provide an extended introduction, reviewing both
the standard-  and the  dilaton-improvement, and briefly discussing how the latter could be relevant in the context of strong interactions.

\subsection{The improved  free scalar field reviewed}
\label{sec:review}

It is well known that in curved space, besides the kinetic term, there is  a second one proportional 
to the Ricci scalar (see \cite{Feynman:1996kb}).\footnote{We use the $(-,-,-)$ sign convention in the notation of 
\cite{Misner:1973prb}.}
%\footnote{
%At the level of a four-derivative kinetic term there exists a unique conformal extension \cite{Riegert:1984kt}.  
%It has been argued that such theories can be ghost free \cite{Bogolyubov:1990kw,Rivelles:2003jd,Bender:2007wu}  and that they have interesting properties for models of cosmology \cite{Boyle:2021jaz,Turok:2022fgq}. The treatment of this case is outside the scope of this paper.} for a massless scalar field 
\begin{equation}
\label{eq:Lag}
\Lag_{\varphi} = \frac{1}{2}  \left( (\partial \varphi)^2 - \xi R \varphi^2 \right) \;.
\end{equation}
This term might  be regarded as a non-minimal coupling to gravity and the specific value of the $\xi$-parameter
is related to the improvement discussed  below.
The corresponding classical energy momentum tensor (EMT), 
defined by the metric variation, reads
\begin{equation}
\label{eq:Tmunuphi}
T_{\mu \nu} =2  \frac{ \de  }{  \de{g^{\mu\nu}}} \int d^4 x \sqrt{-g} \Lag_{\varphi} \Big|_{g_{\mu\nu} = \mink_{\mu\nu} }  = 
 \partial_{\mu} \varphi  \partial_{\nu} \varphi  - \frac{\mink_{\mu\nu}}{2}  ( \partial \varphi)^2 + 
  \xi  (\partial^2 \mink_{\mu\nu}-  \partial_{\mu} \partial_{\nu} )\varphi^2 \;,
\end{equation}
where, for simplicity,  the flat space limit has been assumed with $\mink_{\mu\nu} = \textrm{diag}(1,-1,-1,-1)$.
Taking the trace and using the equation of motion 
for the massless scalar  $\partial^2 \varphi =0$,
one gets
\begin{equation}
\label{eq:TEMT}
\TEMT  = - \dphi  ( \partial \varphi)^2 + \xi (d-1) \partial^2 \varphi^2   = 
(d-1) ( \xi - \xid)    \partial^2 \varphi^2 \;,
\end{equation}
where 
\begin{equation}
\label{eq:dphi}
\xid \equiv  \frac{\dphi}{2(d-1)}\; \stackrel{d = 4} {\longrightarrow} \; \frac{1}{6}  \;, \quad \quad  
\dphi  \equiv \frac{d-2}{2} \;, 
\end{equation}
with $\dphi $ the dimension of the free scalar. Leading order  conformality $\TEMT =0$ is assumed 
 when  $\xi = \xid$ is chosen  and corresponds to the famous improvement. 
Conformality in $d=2$ is automatic in that $\xi_2 =0$ and thus $\xi =0$, 
 where the free scalar field serves as a simple example of a  conformal theory 
see\,\cite{DiFrancesco:1997nk}. 

Historically, the value $\xi_4 = \frac{1}{6}$ was first noted in the context of 
conformal symmetry in general relativity  \cite{DeWitt:1964oba} 
and later  seen as a necessary choice
to obey  the weak equivalence principle  \cite{Chernikov:1968zm}  (or also   \cite{Grib:1995xm}). 
In quantum field theory it was shown that    $\xi = \xid$ leads to a  UV-finite EMT for the free field 
which is a 
necessity, in a renormalisable theory, 
since $T_{\mu \nu}$ is an observable \cite{Callan:1970ze}.  
The authors, Callan, Coleman and Jackiw, referred to the corresponding EMT as
``the improved EMT" which has become standard terminology. 
Similarly  the value $\xi = \xid$  guarantees UV-finiteness of the integrated Casimir energy \cite{DeWitt:1975ys}.

\subsection{The   Goldstone improvement problem}
\label{sec:problems}

 Dolgov and Voloshin  concluded that $\xi = 0$ for Goldstones by applying a soft theorem 
 to gravitational form factors  \cite{Voloshin:1982eb}   (see  also  \cite{Leutwyler:1989tn,Donoghue:1991qv} 
 and \SEC\ref{sec:low}).  For concreteness, the standard 
 chiral spontaneous symmetry breaking  of QCD-like theories, $SU(N_F)_L \times SU(N_F)_R \to SU(N_F)_V$, is assumed and its Goldstones are  
  referred to as pions, see   \cite{Donoghue:1992dd,Scherer:2012xha}.  We may think of the quark condensate 
  $\vev{\bar qq} \neq 0$ breaking \emph{both} chiral and conformal symmetry spontaneously.
  The improvement  problem can be seen  in that a term  as in \eqref{eq:Lag}
    cannot be constructed from the  coset field $U = \exp( i \pi/F_\pi)$ without breaking 
 $SU(N_F)_L \times SU(N_F)_R$-invariance: for example   $\de \Lag \propto R \Tr [U + U^\dagger]$. 
 This  applies equally to abelian Goldstones 
 such as the $\eta'$ or the axion since the shift symmetry forbids writing these terms. 
 Is this a problematic?  Not for the UV divergences as  the EMT requires  extra renormalisation in an effective theory. 
 However, the  point about the   weak equivalence principle remains and a non-vanishing classical trace 
 causes potential issues for   flow theorems  see \APP\ref{app:flow}.

   \subsection{The dilaton can and must be improved}
   \label{sec:SSB}

 The  order parameter of spontaneous scale symmetry breaking  is the dilaton decay constant
 \begin{equation}
 \label{eq:FD}
\matel{0}{T_{\mu\nu}}{\dil(q)} =  
\frac{F_{\dil} }{\dbar}  
(m_{\dil}^2 \mink_{\mu\nu} - q_\mu q_\nu) \;,
\end{equation} 
where we have allowed for a dilaton mass in addition.  The normalisation $\vev{D(\vec{q})|D(\vec{p})} = (2 \pi)^{d-1} E_D \de^{(d-1)}(\vec{p}-\vec{q})$, implies a mass dimension  $[F_D] = \dphi$.  The analogy with the pion decay constant, the order 
parameter of chiral symmetry breaking \cite{Goldstone:1962es,Donoghue:1992dd,Scherer:2012xha}, becomes apparent 
when considering the dilatation current $J^{\dil}_\mu(x) = x^\nu T_{\nu\mu}$ with matrix element $\matel{0}{J^{\dil}_\mu}{\dil(q)} =   i {F_{\dil}} q_\mu  $ by  \eqref{eq:FD} 
upon using $x^\nu \to-i  \partial_{q_\nu}$.
The main point we wish to make is that the dilaton can and must be improved since it is the improvement 
term that realises the fundamental matrix element \eqref{eq:FD} in the effective theory. 
To see this consider the dilaton described by the coset field 
\begin{equation}
\label{eq:Ddef}
 \hat{\chi} \equiv  e^{-\frac{D}{F_D}} \;, \quad 
\chi  \equiv  ( F_D/\dphi)   \hat{\chi}^{\dphi}   \;, \quad
\end{equation}
convenient for the effective theory. They 
transform non-linearly 
\begin{equation}
{D} \to {D} - F_D \al(x)  \quad \Rightarrow \quad 
 \XX \to e^{  \al(x) \dphi } \XX \;, \quad \hat{\XX} \to e^{  \al(x)  } \hat{\XX} \;,
\end{equation}  
under Weyl transformations  ($g = \det(g_{\mu\nu})$)
\begin{equation}
\label{eq:Weyl}
g_{\mu\nu} \to e^{-2 \al(x)} g_{\mu\nu}  \quad \Rightarrow \quad 
\sqrt{-g} \to e^{-d \al(x)} \sqrt{-g}  \;, \quad R \to e^{2\al} R + \ORD(\partial \al) \;.
\end{equation}
The field $\XX$ is convenient since it transforms like the free scalar and thus the improvement 
in \eqref{eq:Lag} simply reads
\begin{equation}
\label{eq:Lchi}
\Lag_{\XX} =  \frac{1}{2} (\partial\XX)^2 - \frac{\xid}{2} \, R \, \XX^{2} \;,
\end{equation}
such that  the sum of the  two (globally Weyl-invariant) terms in \eqref{eq:Lchi} become \emph{locally} Weyl-invariant, see  for example 
\cite{Karananas:2015ioa}.\footnote{In a UV complete theory 
 the $\xid$-term \eqref{eq:Lchi} is to be seen as  a LO  low-energy constant. 
 The same applies to the original  Callan-Coleman-Jackiw improvement  of the free scalar \cite{Callan:1970ze}; not as an effective theory but to ensure the nonrenormalisation  of the energy momentum tensor. 
Corrections  would not change the deep infrared picture  and be partially captured by the running 
 of    $\xid$,  as briefly reviewed   in \APP\ref{app:run}.}
Varying the improvement term with respect to the metric \eqref{eq:Tmunuphi} leads to an expression 
\begin{equation}
\label{eq:TRdef}
  T^R_{\mu\nu} =  \xid (\mink_{\mu\nu} \partial^2 -  \partial_{\mu} \partial_{\nu} )    \XX^{2} \;,
\end{equation}
which realises  the fundamental matrix element
\begin{equation}
\label{eq:TR}
  \matel{0}{  T^R_{\mu\nu}  }{\dil} 
=  \frac{F_{\dil} }{\dbar} ( m_{\dil}^2 \mink_{\mu\nu} -   q_\mu q_\nu)  \;,
\end{equation}
 when  $\xid$ assumes the value in \eqref{eq:dphi}.  
As for the pion theory, this can be seen as the reason for the decay constant $F_D$ appearing in the coset field \eqref{eq:Ddef}.
From a  pedestrian viewpoint the term linear in $T_{\mu\nu}  \supset F_D/(d-1) \partial_\mu \partial_\nu D$ cannot come from 
either a  potential  nor a  kinetic term. 
 The potential term has no derivatives and 
  the kinetic-term derivatives act on two dilatons and thus vanish when evaluated on the vacuum-to-dilaton matrix element.  Thus the need for the improvement term in \eqref{eq:Lchi}.

We end the introduction by discussing the status of  spontaneous  conformal symmetry  breaking.   
It is well known that spontaneous breaking of both, scale and conformal symmetry, 
give rise to a single Goldstone, the dilaton \cite{Isham:1970gz,Coleman:1985rnk,Low:2001bw}.  
That the additional special conformal  symmetry generators   do not lead to extra Goldstones  is a peculiarity of 
spacetime symmetries. Whether or not  scale invariance implies conformal invariance    
  is a matter of debate (see \cite{Nakayama:2013is}  for a review 
and Refs.\cite{Fortin:2012hn,Luty:2012ww,Bzowski:2014qja,Dymarsky:2014zja,Dymarsky:2015jia} for an ever closer understanding). It is noteworthy that the effective field theory (EFT)  itself seems to favour conformality 
for the leading term.\footnote{Weyl-invariance can be seen as the generalisation of conformal invariance to 
curved space. Since we mostly work in flat space Weyl- and conformal-invariance are sometimes used interchangeably.}  We shall see this remains the case when pions are added.  
We treat the dilaton phase purely formally in this paper, but it seems worthwhile to comment on its plausibility.  
It is generally accepted that a massless dilaton describes a spontaneously broken phase of a conformal  theory,  see for example \cite{Schwimmer:2010za}. Examples  are known in
$d=2$ \cite{Semenoff:2018yrt} at finite temperature and  $d=3$ \cite{Bardeen:1983rv} for a non-trivial UVFP.  
It is not settled whether this still holds for a theory  with IR-emergent conformal symmetry  (i.e. when 
the theory  flows into an IR fixed point (IRFP)).  Does the corresponding  Goldstone candidate acquire a mass through the trace anomaly,  similar to the $\eta'$ in QCD, or does it remain massless? It is a question we cannot answer, yet. 
At least the example of a $d=3$  Gross-Neveu-Yukawa theory is  of the massless type and includes a non-trivial flow
\cite{Cresswell-Hogg:2023hdg,Semenoff:2024prf,prep}.  In addition, explorations of a dilaton in the chirally broken phase 
of QCD and ${\cal N}=1$ supersymmetryic gauge theories \cite{Zwicky:2023bzk,Zwicky:2023krx} 
are  consistent with  theoretical and empirical knowledge.\footnote{Other attractive features are that massive hadrons are not in conflict with IR conformality \cite{DelDebbio:2021xwu} (see also \SECs\ref{sec:WG}, \ref{sec:NGFF}) and that 
 it could  ease the explanation of $K \to \pi\pi$ \cite{Crewther:2012wd,Crewther:2015dpa}.} 
This includes the prediction that the scaling dimension of the quark condensate is $\De_{\bar qq} = d-2$ 
which  we will recover as a low-energy consistency relation when examining form factors.   
For the purposes of this paper we will  state  in each section whether the dilaton is assumed to be massless or not.  
Note that massive pions due to explicit breaking will always induce a massive dilaton 
via the dilaton Gell-Mann-Oakes-Renner relation  \cite{Zwicky:2023krx}.

The paper is organised as follows.   In \SEC\ref{sec:main} it is shown how improvement is demonstrated  for non-Goldstone 
and Goldstone systems. 
In \SEC\ref{sec:GFF} the impact of the improvement term is worked out for 
  gravitational form factors of several particle types.  
 The paper ends with summary and conclusions in \SEC\ref{sec:conc}.  
 Appendices \ref{app:earlier} and \ref{app:RG} concern the relation to perturbative models 
 and  renormalisation group matters,  such as the relevance of the improvement/dilaton  for  flow theorems.

 \section{Dilaton Improvement}
 \label{sec:main}
 
We find  it instructive to   demonstrate how the dilaton improves ordinary particles.  A central element is the Weyl-weight 
which can be considered an analogue of the electric charge with regard to dilatations. 
In a UV-complete theory, its value is generally unknown but fixed by the underlying dynamics.
 An exception is the pion whose non-linear representation enforces zero Weyl-weight.  
The results obtained will then be applied to the gravitational form factors in the next section.
  
 \subsection{\RZ{Weyl-gauged scalar and spin-$\frac{1}{2}$ fermion}}
 \label{sec:WG}
  
 We consider a generic scalar  $\scal$ and Dirac fermion $\ferm$ of arbitrary Weyl-weights $\cw_{\scal,\ferm}$ 
\begin{equation}
 \scal  \to e^{\al \cw_\scal} \scal \;, \quad \ferm  \to e^{\al \cw_\ferm} \ferm \;,
\end{equation}
 under Weyl transformation \eqref{eq:Weyl}.  
The question is how to write a locally  Weyl-invariant Lagrangian.   
  The idea, developed in the early 70's \cite{Isham:1970gz,Ellis:1970yd},  is to turn the Weyl symmetry into a local symmetry where the dilaton takes on the role 
 of the gauge field. 
 The key is to introduce the Weyl covariant derivative 
\begin{equation}
 \Wcov_\mu \phi =
 \big( \partial_\mu + (\cw_\phi g_{\mu\nu} + i \Sig_{\mu\nu})  (\partial^\nu \hat{D}  ) \big)  \phi \;, 
 \quad \hat{D} \equiv D/F_D \;,
\end{equation}
where $\Sig_{\mu\nu}$ is the generator of Lorentz transformations and $\phi$ a field of unspecified spin.
 For the scalar and the fermion we have $\Sig_{\mu\nu}|_{\scal} = 0$ and
 $\Sig_{\mu\nu}|_{\Psi}  = 
\frac{i}{4}[\ga_\mu,\ga_\nu]$, yielding 
 \begin{equation}
\Wcov_\mu \scal = (\partial_\mu + \cw_\scal  (\partial_\mu \hat{D}  )) \scal \;,  \quad  
 \slashed{\Delta} \ferm = 
 \big( \slashed{\partial} + (\cw_\ferm - d_\ferm) \slashed{\partial} \hat{D} \big)  \ferm \;,
\end{equation}
where $d_\ferm = \frac{d-1}{2}$ and $ \slashed{a} = \ga_\mu a^\mu$, 
such that 
the field derivatives 
\begin{equation}
\label{eq:W2}
\Wcov_\mu \scal \to e^{ \al(\cw_\scal+1)} \Wcov_\mu \scal  \;, \quad \slashed{\Wcov} \ferm \to e^{ \al \cw_\ferm} \slashed{\Wcov} \ferm 
\;,
\end{equation}
transform covariantly with Weyl-weights $\cw_\scal+1$ and $\cw_\ferm$, respectively.
We may  write the locally Weyl-invariant $\scal$-$\ferm$-dilaton Lagrangian 
\begin{equation}
\label{eq:LphiD}
{\cal L}_{\scal,\ferm,\chi} =  
\Lag_{\XX} +    \frac{1}{2}  \hat{\XX}^\beS  (  (\Wcov \scal)^2 - \hat{\chi}^{2} m_\scal^2  \scal^2  )
 +  \hat{\XX}^\beF \bar{\ferm}  (i \slashed{\Delta} - \hat{\XX} m_\ferm)  \ferm  
 \;,
\end{equation}
with $\Lag_{\XX} $ given in \eqref{eq:Lchi} and 
\begin{equation}
\label{eq:be}
\beS \equiv 2 ( \dphi  - \cw_\scal ) \;, \quad 
 \beF \equiv   2 ( d_\ferm - \cw_\ferm)     \;. \quad  
\end{equation}
The $\hat{\XX}$ prefactors can be seen as conformal compensators, see 
for example \cite{Coleman:1985rnk}.
We have added mass terms assuming that they are not due to explicit symmetry breaking.  We will comment on the latter case 
further below. 
The reason there is no Weyl-covariant derivative for the dilaton is that it vanishes
  $\Wcov_\mu \chi = 0$, which is easily verified.  Hence,   there is not alternative and 
  improving the dilaton is thus mandatory. 
    We will see that for the pion it is not needed either as it is of  Weyl-weight  zero.

\subsubsection*{Energy momentum tensor trace of the scalar}

 We focus on the scalar, as the fermion works in a similar manner.   
 In order  to simplify the  equations we shall temporarily assume $d=4$. 
 The EMT is obtained by 
 metric-variation  and assuming a flat spacetime we get
  \begin{equation}
  \label{eq:EMTphi}
  T_{\mu\nu} =  \hat{\XX}^\beS \Wcov_\mu \scal \Wcov_\nu \scal 
     +  \partial_\mu \XX  \partial_\nu \XX
   - \mink_{\mu\nu} \Lag_{\scal\text{-}\chi}  + \TImp_{\mu\nu}\;,
   \end{equation} 
   with $\TImp_{\mu\nu}$ given in \eqref{eq:TRdef}.
To investigate its trace one must take into account the equation of motion. For 
 $\scal$  and the dilaton they read
\begin{equation}
\label{eq:phieom}
{( \partial^2 + \hat{\chi}^2 m_\scal^2 +  [ \cw_\scal  \partial^2 \hat{D} - \cw_\scal (\cw_\scal+\beS)(\partial \hat{D})^2 - \beS \partial{\hat{D}} \cdot \partial] ) \scal   
=0} \;,
\end{equation}
and 
\begin{alignat}{2}
& \hat{\XX}^{-\beS } \chi \partial^2 \chi &\;=\;&  \partial^\mu( ( \partial_\mu \scal) \scal) +  \partial^\mu(( \partial_\mu D) 
\scal^2) -   \cw_\scal \beS \scal \partial \hat{D} \! \cdot \! \partial \scal -
\nonumber \\[0.1cm]  
&   &\;\phantom{=}\;&     \cw_\scal^2 \beS (\partial \hat{D})^2 \scal^2 
- (1+ \frac{\beS }{2}) \hat{\chi}^2 m_\scal^2 \scal^2 + \frac{\beS }{2} (\Wcov \scal)^2 \;,
\end{alignat}
respectively.  Using 
\begin{equation}
\frac{1}{2} (\Wcov \scal)^2  = \frac{1}{2} (\partial \scal)^2  +  \cw_\scal \,  \scal (\partial{D} \cdot \partial \scal) + \frac{1}{2} \cw_\scal^2 \,
 (\partial {D})^2 \scal^2 \;,
\end{equation}
and \eqref{eq:be}, 
we may combine them into a compact   equation   
\begin{equation}
\label{eq:chieom}
{\chi \partial^2 \chi =   \hat{\XX}^\beS (\Wcov \scal)^2 - 2 \hat{\chi}^{\beS+2} m_\scal^2 \scal^2}  \;,
\end{equation}
 which is Weyl-invariant up to the kinetic term.  Manifest invariance can be restored  by 
adding curvature terms of the $R \XX^{d-2}$-type.  
Since we work in flat space this subtlety is  of no relevance and thus 
ignored. 
We are  now in a position to verify conformality by contracting \eqref{eq:EMTphi}
\begin{alignat}{2}
\label{eq:TEMTphi}
 & \TEMT &\;=\;&     3 \xif  \partial^2 \XX^2   -  (\partial \chi)^2 - \hat{\XX}^\beS (\Wcov \scal)^2  +  2 \hat{\chi}^{\beS+2} m_\scal^2 \scal^2   \nonumber \\[0.1cm]  
  &  &\;\stackrel{\eqref{eq:chieom}}{=}\;&  
    3 \xif  \partial^2 \XX^2   -  (\partial  \XX )^2   -  \XX  \partial^2 \XX  
  =     ( 6 \xif -1) \big\{  \XX \partial^2 \XX +   (\partial  \XX )^2  \big\} 
   \stackrel{\xif =1/6}{\longrightarrow}
 0   \;,
\end{alignat} 
using the equation of motion and substitution of 
the appropriate improvement parameter.  
This is the expected result as the Weyl-covariant derivative renders 
\eqref{eq:LphiD}  locally Weyl-invariant.

 The vanishing of the trace of the EMT for a massive scalar particle at LO realises the scenario of “massive hadrons in a conformal phase” \cite{DelDebbio:2021xwu}. In that reference, it was shown that $\matel{\scal(p)}{ \TEMT}{\scal(p)} =0$
with $m_\varphi \neq 0$, for zero momentum transfer and $\cw_\varphi = 1$, using only the LSZ reduction procedure. In this case, the Weyl compensator on the mass term proved sufficient, as the Weyl-covariant derivative affects only the $\mathcal{O}(q^2)$ behaviour.
We believe that incorporating the improvement term  together with the Goldstone matrix element \eqref{eq:TR},  into the effective  theory framework, clarifies the derivation. 
Nonetheless, if conformal symmetry is only emergent, we anticipate corrections at NLO.

It is curious to have the two alternatives for the scalar: Weyl-gauging or Ricci-gauging (term used in \cite{Iorio:1996ad} for the 
standard Callan-Coleman-Jackiw improvement).   The general conditions of when Ricci-gauging provides an alternative to Weyl-gauging 
has been worked out in \cite{Iorio:1996ad}. We do stress that the two are generally dynamically inequivalent 
\emph{and} that the  dilaton is a specific interpretation of the Weyl gauge field.

 \subsection{The pion-dilaton system}
 \label{sec:piD}
 
 Whereas  the Weyl-weight is a  non-trivial  dynamical quantity in general it is known for the pions. 
 Due to their non-linear representation it can be deduced from the conformal algebra  \cite{Ellis:1970yd} 
 and its weight is found to be zero.  
Intuitively this expresses the fact that pions are generalised angles, which are not expected to be sensitive to dilatations. The verification of the tracelessness is therefore quasi-identical for this system with $\cw_\pi =0$.
Nevertheless,  we present it while generalising  to 
a curved $d$-dimensional space to ensure an  element of novelty. 

Neglecting terms in the EFT, suppressed by the cutoff 
$\Lambda \approx 4 \pi   F_{D,\pi}$, 
the LO Lagrangian   reads 
  \begin{equation}
  \label{eq:LLO}
\Lag_{d \chi \text{PT}} =  \Lagkind   -  \frac{\xid}{2}   R \, \XX^{2}  - \potd(\XX)   \;,
 \end{equation}
 with kinetic term
 \begin{equation}
 \label{eq:Lkin}
  \Lagkind =  \Lagkind^\pi +  \Lagkind^D =       \frac{ F_\pi^2 }{4}  \hat{\XX}^{d-2}  \Tr [ \nabla^\mu  U \nabla_\mu  U^{\dagger}] +   \frac{1}{2}(\nabla\XX)^2 \;,
 \end{equation}
 and $\potd$ denotes  a generic potential $\potd$  to track  symmetry breaking ($ \TEMT = \ORD(\potd)$). Here,  $\partial \to \nabla$ is the diffeomorphism covariant derivative ($\nabla_\al g_{\mu\nu}=0$)  
 and further note the alternative form $ \hat{ \XX}^{d-2} (\nabla {D} )^2 =  (\nabla \XX)^2$ for the kinetic term.
  Crucially, the prefactor $\hat{\XX}^{d-2}$ in front of $ \Tr [ \nabla^\mu  U \nabla_\mu  U^{\dagger}] $ signals that  pions are of zero Weyl-weight (as does $\Wcov U \to  \nabla U$). 
  We shall refer to this EFT as 
  dilaton chiral perturbation theory (\chiPT\!\!).
 
 The dilaton equation of motion  reads
 \begin{equation}
 \label{eq:Veomd}
\XX  \nabla^2 \XX = (d-2) (  \Lagkind^\pi  +  \LagImp_d)  -   \partial_{\ln \XX} \potd \;.
 \end{equation} 
The EMT is given by
 \begin{equation}
  T_{\mu\nu} = 
     \frac{F_\pi^2}{2}  \hat{\XX}^{d-2}   \Tr [ \nabla_\mu  U \nabla_\nu  U^{\dagger}] + 
     \nabla_\mu \XX  \nabla_\nu \XX   - g_{\mu\nu}( \Lagkind- \potd )   + \TImp_{\mu\nu} \;,
   \end{equation}
where the  improvement  reads
  \begin{equation}
  \label{eq:Timpd}
   \TImp_{\mu\nu} =   \xid  \big( 2 G_{\mu\nu}  + (g_{\mu\nu} \nabla^2 -  \nabla_{\mu} \nabla_{\nu} )    \big)  \XX^{2}  \;,
  \end{equation}
  with $G_{\mu\nu} = R_{\mu\nu} - \frac{1}{2} R g_{\mu\nu}$ the conserved Einstein tensor, which implies that   
  transversality 
holds separately:    $\nabla^\mu \TImp_{\mu\nu}  = 0$.
 Taking the trace and using the  equation of motion  we obtain
\begin{alignat}{2}
\label{eq:TEMTmain}
 & \TEMT &\;=\;&     (d-1)  \xid  \nabla^2 \XX^{2}     -  (d-2)(   \Lagkind^\pi  +  \LagImp_d + 
 \Lagkind^D )     
 + d \potd \nonumber \\[0.1cm]  
  &  &\;\stackrel{\eqref{eq:Veomd}}{=}\;&   (d-1)  \xid  \nabla^2 \XX^{2}   - \dphi   (\nabla  \XX )^2    - \XX   \nabla^2 \XX   
    + {\cal F}_d(\potd)  \nonumber \\[0.1cm]
  &  &\;=\;&     ( 2 (d-1) \xid   -\dphi)   \big\{  \XX \nabla^2 \XX +  (\nabla  \XX )^2  \big\} + {\cal F}_d(\potd) \;,
\end{alignat} 
where  ${\cal F}_d(\potd) = (d   - \partial_{\ln \XX}) \potd $. 
  Crucially, 
  \begin{equation}
  \label{eq:tadpole0}
 \TEMT =    {\cal F}_d(\potd) =     F_D^2 m_D^2 \hat{D}  +  \ORD(\hat{D}^2) \;,
 \end{equation}
 holds provided $
 \xid $ is given as in \eqref{eq:dphi}.  The presence of the linear term (tadpole) is 
 characteristic for the divergence of the  Goldstone current. It  
 will play an important role for the form factors.

 Whereas the linear term is universal the quadratic term in $\TEMT$  is model-dependent (on the cubic term in the potential).
 Hence the question of whether  anything generic  can be said about the potential? 
 If scale symmetry is   unbroken then the only scale invariant term is
 $\potd \propto \XX^d$ which is not permitted since there is no local minimum and thus $V_d =0$. 
 This necessitates the introduction of an additional term 
 due to some operator breaking scale symmetry \cite{Zumino:1970tu}.
 For the case of a  single operator ${\cal O}$, 
 we refer the reader to  \cite{Zwicky:2023krx}, where it is found that $\De_{\cal O} = d-2$ by soft-theorem 
 arguments.\footnote{In the case of the gauge theory the only permissible operator seems to be the quark bilinear $\bar qq$ which requires an explicit  fermion mass term, 
 and its scaling dimension is found to be $\De_{\bar qq} =d-2$, 
 satisfying the constraint.  
The gluon field strength squared cannot take this role since its scaling dimension is $\De_{G^2} =d.$ \cite{Zwicky:2023krx}.}  The generic case is more involved and less understood.

  \section{Gravitational Form Factors  }
 \label{sec:GFF}

 The improvement term does not affect scattering amplitudes in flat space  
 but it alters gravitational form factors since it changes the EMT.  
 Gravitational form factors are one-particle transition matrix elements of the EMT; 
 the analogue of the electromagnetic pion form factor with regard to gravity for example 
 $\matel{\pi(p')}{T_{\mu\nu}}{\pi(p)}$.
 They  are widely studied for the  nucleon, as reviewed in \cite{Polyakov:2018zvc}, 
 as they are accessible in hard-exclusive processes 
 \cite{Burkert:2018bqq,Duran:2022xag}  and they are related to the mass decomposition of 
 the nucleon \cite{Lorce:2017xzd,Liu:2021gco,Ji:2021qgo,Ji:2021mtz}.
We focus on the cases of the previous section, the scalar 
 $\scal $  and   the fermion $\ferm$ in  flat-space where the former is relevant for the dilaton and pion as well. 
 We introduce the shorthand  
 \begin{equation}
\label{eq:RZ}
\GaC_{\mu\nu}^\scalg \equiv 
 \matel{\scalg(p')}{T_{\mu\nu}(0)}{\scalg(p)}  \;, \quad \GaC_{\mu\nu}^\ferm   \equiv 
    \matel{\ferm(p',s')}{T_{\mu\nu}(0)}{\ferm(p,s)} \;,  
\end{equation}
 to parametrise the form factors   by
\begin{alignat}{2}
  \label{eq:GF12}
  & \GaC_{\mu\nu}^\scalg &\;=\;& 
  2 {\cal P}_\mu {\cal P}_\nu \gone^\scalg (q^2) +
\frac{1}{2} (q_\mu q_\nu-  q^2 \mink _{\mu\nu} ) {\gtwo}^\scalg (q^2)  \;,  \\[0.1cm] 
  &\GaC_{\mu\nu}^\ferm    &\;=\;& \frac{1}{2 m_{\ferm}} 
   \bar u(p') \left(  2  {\cal P}_\mu {\cal P}_\nu  \gone^\ferm (q^2)  +  
   2  {i  {\cal P}_{\{ \mu } \sig_{q \nu\} }}\,  \gJ^\ferm(q^2)  +   
  \frac{1}{2} (q_\mu q_\nu-  q^2 \mink _{\mu\nu} ) {\gtwo}^\ferm (q^2)    \right) u(p) \;,
    \nonumber
\end{alignat}
where $\bar u(p) u(p) = 2m_\ferm$, 
$\sig_{q \mu } = \sig_{\nu\mu}q^\nu$, 
$\sig_{\mu\nu} \equiv  \frac{i}{2} [ \ga_\mu,\ga_\nu] $ (equal to $2 \Sig_{\mu\nu}$ the spin generator used previously), 
${\{\mu , \nu\} } = 
 {\mu}{\nu }- {\nu}{\mu }$  and
\begin{equation}
q \equiv p'-p \;, \quad  {\cal P} \equiv 
\frac{1}{2}(p + p') \;,
\end{equation}
are  momentum transfer and  average respectively.  Translational invariance implies  $q^\mu \GaC_{\mu\nu}   =0$, 
thereby  
limiting  the number of form factors.\footnote{In comparison with  \cite{DelDebbio:2021xwu}, $G_1 \to \gone$ and 
$m^2_\scalg/q^2 G_2 \to \frac{1}{2} \gtwo$, as the mass proves, 
 inconvenient for the massless case and we changed the sign convention of $q$ to harmonise with the literature  \cite{Hudson:2016gnq}.} The link to the total momentum  constrains  the $A(q^2)$ form factors at zero momentum transfer
\begin{equation}
\label{eq:g1}
P_\mu = \int d^3 x \, T_{0\mu} \quad \Rightarrow  \quad \gone(0)  = 1 \;,
\end{equation}
for state normalisation $\vev{\varphi(p')|\varphi(p)} =
 (2 \pi)^{d-1} 2 E_\varphi \de^{(d-1)}(\vec{p} - \vec{p}\,')$.  The form factor $J$ is related to the gravitational 
 anomalous magnetic moment which vanishes and sets the additional constraint $J^\ferm(0) = \frac{1}{2}$  \cite{Hudson:2016gnq}.\footnote{A result due to the universality of gravity as the analogue of the fermionic contribution to the electron magnetic anomalous moment is cancelled by a bosonic contribution of the photon which couples to gravity.}
For later convenience we define the trace of the form factors, 
$\gzero^\scalg  \equiv \GaC_{\mu}^{\scalg\, \mu}$ in \eqref{eq:GF12},  such that 
\begin{alignat}{2}
\label{eq:g0}
& \gzero^\scalg(q^2) &\;=\;&
  2 m_\scalg^2 \,  \gone^\scalg(q^2) - \frac{q^2}{2} \,( \gone^\scalg(q^2) + (d-1) {\gtwo }^\scalg(q^2)) 
  \;, \quad  \nonumber \\[0.1cm]
& \gzero^\ferm(q^2) &\;=\;&
  2 m_\ferm^2 \,  \gone^\ferm(q^2) - \frac{q^2}{2} \,( \gone^\ferm(q^2)  -   2 \gJ^\ferm(q^2)+ (d-1) {\gtwo }^\ferm(q^2)) 
  \;.
\end{alignat} 
Assuming that $ {\gtwo}$ has no pole, \EQ\eqref{eq:g1}  implies 
the textbook formula  $ \gzero^\scalg(0) = 2 m_\scalg^2$ \cite{Donoghue:1992dd}.
Crucially, this is no longer true when the dilaton is massless, as then the  $2 m_\scalg^2$ is cancelled by the dilaton pole 
\begin{equation}
\label{eq:conf}
{\gtwo}^\scalg(q^2)|_{m_D=0} =  \frac{4 }{d-1} \frac{m_\scalg^2}{q^2} + \de {\gtwo}^\scalg(q^2) \;,
\end{equation}
to yield a traceless EMT $\gzero^\scalg(q^2) =0$. This is  \emph{the mechanism} by which  \eqref{eq:TEMTphi}
is realised.  One of the main goals of this section is to understand this  in more detail. 

Before proceeding, let us briefly discuss the normalisation of the form factors. 
\RZ{First, one might be concerned  that the normalisation \eqref{eq:g1} is altered by the dilaton pole.  
However,  
restoring 
the $x$-dependence $\GaC_{\mu\nu}^\scalg \to e^{i q x} \GaC_{\mu\nu}^\scalg$ in \eqref{eq:RZ}, reveals that there is a preferred direction: $\vec{q} = 0$ or 
$q = (\sqrt{q^2},\vec{0})$; as otherwise the  relation in \eqref{eq:g1} won't hold. 
In this limit, it is straightforward  to see that the normalisation remains unaffected 
and thus $A(0)=1$ holds regardless. 
Second, in the case of a spontaneously broken CFT which in particular implies 
a massless dilaton and $\gzero(q^2) = 0$, \eqref{eq:g0} then implies 
($A(0)=1$ and   $J(0) = 1/2$) 
\begin{alignat}{2}
\label{eq:Ap}
& A^{'\scalg}(0)|_{m_D=0}  &\;=\;&  - \frac{1}{4 m^2_\scalg} ( 1 +  (d-1)  \de {\gtwo}^\scalg(0))  \;, \nonumber \\[0.1cm] 
& A^{'\ferm}(0)|_{m_D=0}  &\;=\;& -  \frac{ (d-1)}{4 m^2_\ferm}   \de {\gtwo}^\ferm(0) \;,
\end{alignat}
with $\de {\gtwo}$  the non-dilaton-part as per above.  
In analogy to the  electromagnetic pion form factor, one can 
compute the mass and energy radii from the slope  \cite{Polyakov:2018zvc}.  
However, it is not clear how to make use of these relations in our LO-context, since $A^{'\scalg}(0)$ gets contributions at NLO only, analogous   to the  pion form-factor case.

\subsection{Scalar and fermion form factors }
\label{sec:NGFF}

As stated above, we wish to investigate the dilaton decoupling in the form factors to better understand  the 
seemingly discontinuous transition from the  massless to the massive case
\begin{equation}
 \label{eq:lthm2}
 \gzero^\scal(0)  = \begin{thincases}
0   &  m_D = 0  \\[-0.1cm]
2 m_\scal^2 &  m_D \neq 0
\end{thincases} \;.
 \end{equation}
Recall that the first equation is a  result of our finding in \SEC\ref{sec:WG} 
and the second is the standard textbook formula  \cite{Donoghue:1992dd}. 
The transition  can be studied by expanding in small and large
 $q^2/m_D^2$
 \begin{equation}
 \label{eq:lthm2}
 \gzero^\scal(q^2)  = 
 \begin{thincases}
\ORD(m_D^2/q^2)    &  q^2 \gg m_D^2  \\[-0.1cm]
2 m_\scal^2(1  + \ORD(q^2/m_{D,\scal}^2)) &  q^2 \ll m_D^2
\end{thincases} \;.
 \end{equation}
  The key is to 
 study the tadpole diagram (see \FIG\ref{fig:dia}), for which  we must determine the effective coupling 
\begin{equation}
\label{eq:deLeff}
\de \Lageff = \frac{1}{2} g_{D\scal\scal} D \scal^2 \;,
\end{equation}
from the Lagrangian \eqref{eq:LphiD}.  Using the kinematics, $p' = p + q$, $p^2 = p'^2 = m_\scal^2$, which leads to 
\begin{equation}
\label{eq:derived}
p' \cdot p = m_\scal^2 - \frac{1}{2} q^2 \;, \quad  p' \cdot q  =  \frac{1}{2} q^2 \;, \quad p\cdot q =  - \frac{1}{2}q^2  \;,
\end{equation}
 we find  (e.g. $(\partial \scal  )^2 
\to  (m_\scal^2 - \frac{1}{2} q^2) \scal^2$)
\begin{equation}
\label{eq:gDphiphi}
g_{D \scalg\scalg}(q^2)  = \frac{1}{F_D} ( [q^2 \cw_{\scalg}  - \beS (m_\scalg^2- \frac{1}{2}q^2)  ]  
+ (\beS+2) {m}_{\scalg}^2  )  = 
\frac{2  {m}_{\scalg}^2   +  \dphi q^2}{F_D} \;,
\end{equation}
where the term in square brackets is from the kinetic term with Weyl-covariant derivative and the second  one from 
the mass term. It is noteworthy that the Weyl-weight drops out of the expression,
We note that the  $q^2$-term is   additional to the expression  in \cite{DelDebbio:2021xwu} 
since it is  due to  Weyl-gauging.
Remarkably, the undetermined Weyl-weight  $\cw_\scal$ cancels in this expression also in the $q^2$-term.  
The tadpole contribution  then reads 
 \begin{equation}
 \label{eq:tadpole}
 \matel{\scal(p')}{ D}{\scal(p)} =  \frac{g_{D\scal\scal}(q^2) }{m_D^2-q^2}  \;.
  \end{equation} 
  Evaluating $\matel{\scal}{T_{\mu\nu}}{\scal} $ for  \eqref{eq:EMTphi}, by using \eqref{eq:tadpole} yields
\begin{equation}
\label{eq:druck}
{\gtwo}^\scal(q^2) =   \frac{4 \xid}{\dphi}  \frac{ F_D \, g_{D\scal\scal}(q^2)}{q^2-m_D^2} - 1 \;.
\end{equation}
Note that the minus one   comes from the kinetic term in $T_{\mu\nu}$ and is standard for the free scalar. 
Combining \eqref{eq:gDphiphi} and  \eqref{eq:druck}, we get the LO formulae 
 \begin{equation}
 \label{eq:fPhiLO}
 \gone^\scal(q^2) = 1 \;, \quad {\gtwo}^\scal(q^2) = \frac{2}{d-1}  \frac{ \dphi q^2 + 2 m_\scal^2 }{q^2-m_D^2} - 1 \;.
 \end{equation} 
 Before studying the quasi-massless and the dilaton-decoupling regime we derive the 
 spin one-half form factors. 
  The effective Lagrangian  contains  three terms 
\begin{equation}
\label{eq:Leffferm}
\Lageff =  {g}^{(1)}_{D\ferm\ferm} \, \bar{\ferm} i ( \slashed{\partial} \hat{D} ) \ferm + 
 {g}^{(2)}_{D\ferm\ferm}\, \hat{D} \bar{\ferm} i ( \slashed{\partial} \ferm )   + 
  {g}^{(3)}_{D\ferm\ferm} \, m_\ferm \hat{D}  \bar{\ferm}  \ferm   \;.
\end{equation}
For on-shell fermions  we may write even more condensed 
\begin{equation}
\label{eq:deLeff}
\de \Lageff =  g_{D\ferm\ferm}\,  D \bar{\ferm}  \ferm   \;,  
\end{equation}
with contributions 
\begin{equation}
\label{eq:gDff}
{g}_{D\ferm\ferm} =\frac{1}{F_D}(0 -m_\ferm \beF + m_\ferm (\beF +1)) = \frac{m_\ferm}{F_D} \;,
\end{equation}
in order of the effective Lagrangian  \eqref{eq:Leffferm}.   The analogous formula of \eqref{eq:druck} 
for the fermions reads 
\begin{equation}
\label{eq:druck2}
{\gtwo}^\ferm(q^2) =    \frac{4 \xid}{\dphi}   \frac{2 m_\ferm F_D \, g_{D\ferm\ferm}}{q^2-m_D^2}  \;,
\end{equation}
without the constant factor characteristic for bosons.
Proceeding as before  we get the LO  fermion expressions 
 \begin{equation}
 \label{eq:ffermLO}
 \gone^\ferm(q^2) = 1 \;, \quad  \gJ^\ferm(q^2) = \frac{1}{2} \;,\quad 
 {\gtwo}^\ferm(q^2) = \frac{2}{d-1}  \frac{ 2 m_\ferm^2 }{q^2-m_D^2}   \;,
 \end{equation} 
 upon using the Gordon identity $2m_\ferm \bar u(p')\ga^\mu  u(p) = u(p') ( 2 {\cal P}^\mu  + i \sig^{\mu\nu}q_\nu)  u(p) $ for  the kinetic part of the EMT.  
The effect of soft-breaking for the scalar and the fermion form factors are deferred to 
\APP\ref{app:soft}. 
     In order to gain some intuition into the dynamics we consider the following two regimes, 
  for the scalar form factors:
  \begin{itemize}
 \item \emph{The quasi-massless dilaton regime:}  expanding in $m_D^2/q^2 \ll 1$
 \begin{equation}
 {\gtwo}^\scal(q^2) =   \frac{4}{d-1} \frac{m_\scal^2}{q^2}  -  \frac{1}{d-1} + 
  \ORD(m_D^2/q^2) \;.
 \end{equation}
Inserting  into \eqref{eq:g0} one verifies 
\begin{equation}
\matel{\scal(p')} {\TEMT}{\scal(p)} =  \ORD(m_D^2) \;,
\end{equation}
as a consistency check of  \eqref{eq:TEMTphi} or \eqref{eq:lthm2}.
\item \emph{The   dilaton-decoupling regime:}  expanding in $q^2/m_D^2 \ll 1$ 
 \begin{equation}
  {\gtwo}^\scal(q^2) =  -\frac{4}{d-1}  \frac{   m_\scal^2 }{m_D^2} -  1   +
 \ORD( q^2/m_D^2) \;.
 \end{equation}
 Inserting we find 
 \begin{equation}
 \matel{\scal(p')} {\TEMT}{\scal(p)}   = 2 m_\scal^2(1  + \ORD(q^2/m_{D,\scal}^2)) \;,
 \end{equation} 
 the standard textbook formula \eqref{eq:lthm2} for when there is no dilaton. 
   \end{itemize}
  Besides the LO results for the scalar  \eqref{eq:fPhiLO}  and the fermion \eqref{eq:ffermLO},
   the understanding  of \eqref{eq:lthm2} consists the main result of this section.

 \begin{centering}
\begin{figure}[h!]
\includegraphics[width=1.0\linewidth]{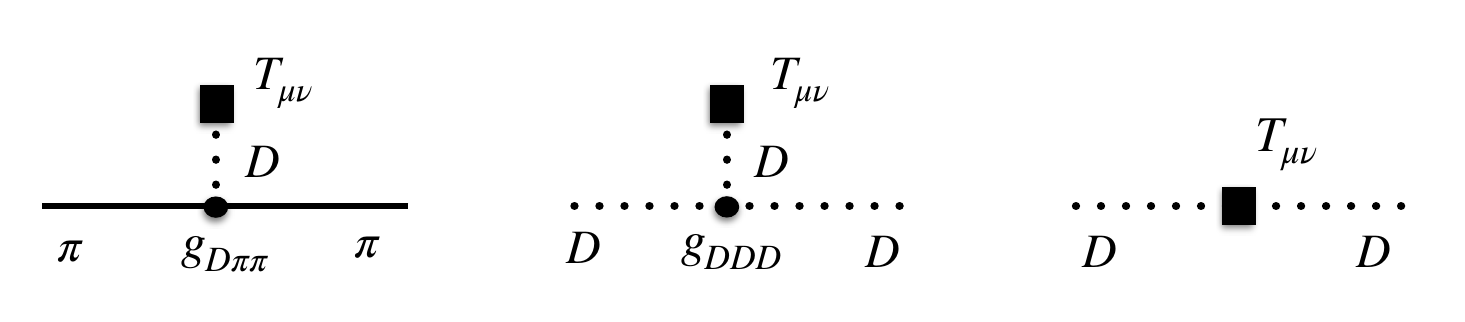}  
\caption{\small  Goldstone gravitational  form factors. (left) pion form factor with $g_{D\pi\pi}$ interaction as in \eqref{eq:gDpipi}
  (centre) analogous contribution to the dilaton  form factor with $g_{DDD}$ as per \eqref{eq:gDXX} 
   (right) contact  interaction term due to the  improvement term. In the sum these last two terms fulfil 	the  constraint in \EQ\eqref{eq:g20} for the dilaton form factor.  Analogous diagrams exist for the non-Goldstone scalar $\scal$ and the fermion $\ferm$.}
\label{fig:dia}
\end{figure}
\end{centering}

  \subsection{A soft-pion theorem}
  \label{sec:low}

Let us specialise to the case where the scalar particle is the pion, $\scalg \to \pi^c$ with flavour index $c$, representing the Goldstone of an internally broken global symmetry. 
The crucial ingredient is the standard soft-pion theorem. 
Let  $Q_5^c$ be the generator of chiral transformation,   
then the soft-pion theorem reads   
\begin{equation}
\label{eq:soft}
 \matel{\pi^c(p')}{T_{\mu\nu}(0)}{\pi^c(p)}  = - \frac{i}{F_\pi} \matel{0}{ \, [Q^c_5,T_{\mu\nu} ] \,}{\pi^c(p)}  +  \lim_{p' \to 0} i p' \! \cdot \! R^{cc} \;,
\end{equation}
 in the limit  $p' \to 0$ with  
\begin{equation}
\label{eq:R}
R^{bc}_\al=    - \frac{i}{F_\pi}\int d^dx\, e^{i p'\cdot x} \matel{ 0}{T J^c_{5\al}(x)T_{\mu\nu} (0)}{\pi^b(p)} \;,
\end{equation}
the remainder term (see, for example \cite{Donoghue:1992dd}). 
Its role is subtle as it vanishes in most cases, but in some cases (e.g. with a massless dilaton) it becomes crucial.
We wish to derive a LO constraint of the following form 
\begin{equation}
\gzero(q^2)|_{\text{LO}} = a + b q^2  + \ORD(q^4) \;.
\end{equation}
In what follows it is important to distinguish the case where the pion is massless or massive. 
The crucial bit is played by the remainder. 
If both the pion and the dilaton are massless, then we know that for the improved system 
$\gzero|_{\text{LO}} = 0$  \eqref{eq:TEMTmain}. For explicit symmetry breaking the pion is massive and so is the dilaton. 
By \eqref{eq:g0} we know that  in this case $a = 2 m_\pi^2$, so we may set the pion mass to zero but 
retain that the remainder \eqref{eq:R} is  zero.
Hence, with  
$[Q_5^a,T_{\mu\nu}] =0$ and  $R^{bc}_\mu=0$ \eqref{eq:soft} we get   
   \begin{equation}
 \matel{\pi^c(p')}{T_{\mu\nu}(0)}{\pi^c(p)}   \to 0 \;, \quad  \text{for} \quad p \to 0 \quad \text{or} \quad  p' \to 0 \;,
  \end{equation}
  and with \eqref{eq:RZ} the constraint 
  \begin{equation}
  \label{eq:softg}
  \gone^\pi(0)  + {\gtwo}^\pi(0) =0 \quad \Rightarrow \quad {\gtwo}^\pi (0) = -1\;,
  \end{equation}
follows upon further using \eqref{eq:g1}.
 We may now compute the trace, restoring the mass of the pion as of above, to obtain 
  \begin{equation}
   \label{eq:lthm} 
 \gzero^\pi(q^2)  =  \begin{thincases} 
 0 + \text{NLO}   &  m_{\pi,D}  =0  \quad   {\gtwo}^\pi(0) = -\frac{1}{d-1}    \\[-0.1cm]
 2 m_\pi^2 + \dphi q^2 + \text{NLO}
   &   m_{\pi,D} \neq 0  \quad  {\gtwo}^\pi(0) = - 1 
      \end{thincases}  \;. 
 \end{equation}
 NLO terms are of the form $m_\pi^4,m_\pi^2 q^2,q^4/(16 \pi F_\pi^2)^2$. 
  The second low-energy theorem is derived in \cite{Novikov:1980fa} for $m_\pi=0$ and used in \cite{Donoghue:1991qv} for  $m_\pi \neq 0$.\footnote{The NLO corrections in the second case have been computed in  chiral perturbation theory in 
  \cite{Donoghue:1991qv}  
 and are of the standard type. The translation of form factor basis reads $\gone^\pi = \theta^\pi_2$, ${\gtwo}^\pi = - \theta_1^\pi$, 
 and $\gzero^\pi = \theta_0^\pi$.}
 It is also important in extracting $F_D$ in the context of the $\sigma$-meson in QCD
 (see \cite{Moussallam:2011zg} and  \cite{Hoferichter:2023mgy}). 
 In what follows we will verify them explicitly in both cases.   
 We will clarify at the end of the next section   whether the soft theorem \eqref{eq:lthm}, second line, applies to the dilaton.

\subsection{Pseudo Goldstones (pion and dilaton)}
\label{sec:MD}

We will assume that there is explicit scale symmetry breaking.  The introduction of mass 
will determine the couplings 
\begin{equation}
\label{eq:geff}
\de \Lageff = \frac{1}{2} g_{D \pi\pi} D \pi^2 +  \frac{1}{3!} g_{D DD} D^3 \;,
\end{equation}
which enter the tadpole diagrams through the linear $D$-term in ${\cal F}_d$
 \begin{equation}
 \label{eq:TEMTuse}
 \TEMT =     m_\pi^2 \pi^2  + F_D^2 m_D^2 \hat{D}  +  \ORD(\hat{D}^2,\pi^4) \;.
 \end{equation}
In the case of the pion 
this is very transparent. As is well known the pion  acquires a mass through the quark mass which is a source of 
explicit breaking. In the EFT the mass term follows from a spurion analysis in standard $\chi$PT.  
Since the mass term generates 
the  quark condensate  $\vev{\bar qq}$, by differentiation of the 
partition function of the microscopic theory, it is then clear that  
we must add the following term to the LO Lagrangian 
\begin{equation}
\label{eq:Lm}
\Lag_{m_\pi} =  
-\frac{1}{2} \vev{\bar qq}   \Tr[{\cal M}U +U^\dagger {\cal M}^\dagger]  \hat{\XX}^{\Dechi} \;,
\end{equation} 
with ${\cal M} = \text{diag}(m_q,\dots ,m_q)$ such that $\Lag_{m_\pi} = -\frac{1}{2} m_\pi^2 \pi^2 + \dots $ emerges 
upon using the Gell-Mann-Oakes-Renner relation $m_\pi^2 F_\pi^2 = - 2 m_q \vev{\bar qq}$.  
Taking into account  the kinetic term  \eqref{eq:Lkin} and \eqref{eq:Lm}, we may 
obtain the dilaton-two-pion interactions 
 \begin{equation}
  \label{eq:LDpipi}
 {\cal L}_{D\pi\pi}  =  \frac{1}{2}  ( \Dechi \, \hat{D} \, m_\pi^2  \pi^2  - 2 \dphi \,   \hat{D} \, ( \partial \pi  )^2)  \;.
 \end{equation}
For the dilaton, the situation is less clear, since it depends on the cubic term in the potential  $V \supset  - \frac{c_3}{3!} D^3$
(see discussion in \SEC\ref{sec:piD}). Leaving   $c_3$ unspecified at first, we get\footnote{\label{foot:VDe}
In the case one associates the dilaton mass with a single operator ${\cal O}$ in the trace of the EMT 
one has 
\begin{equation}
\label{eq:V}
 V_{\De} (\hat{\chi}) =   \frac{m_D^2 F_D^2 }{ \De -d} \left( \frac{1}{\De} \hat{\chi}^\De - \frac{1}{d}  \hat{ \chi}^d \right)   =   
    \text{const}  +    m_D^2 F_D^2 \big( \frac{1}{2} \hat{D}^2 -  \frac{d+\De}{3!} \hat{D}^3  
+ 
\ORD(D^4)\big) \;, 
\end{equation} 
that is  $c_3 =  (d+\De)$ and we refer the reader to \cite{Zwicky:2023bzk} for further discussion and earlier references.} 
  \begin{equation}
 \label{eq:LDDD}
 {\cal L}_{DDD}  = - \frac{D}{F_D} \dphi( \partial D)^2  +  m_D^2 F_D^2 \frac{c_3}{3!} \hat{D}^3 \;.
 \end{equation} 
From \eqref{eq:LDpipi} and \eqref{eq:LDDD} we then obtain  the effective vertices as before 
 \begin{alignat}{4}
 \label{eq:gDpipi}
 &g_{D\pi\pi}(q^2)  &\;=\;&  \frac{1}{F_D} (   (\Dechi &\;-\;&  2 \dphi) m_\pi^2 &\;+\;&  \dphi q^2 ) \;,  \\[0.1cm] 
  \label{eq:gDXX}
   & g_{DDD}(q^2)  &\;=\;&  \frac{1}{F_D}  (     (c_3 &\;-\;& 2 \dphi) m_D^2 &\;-\;&  \dphi  q^2  )   \;,
 \end{alignat}
 where the former is
  well known in the literature (see \cite{Ellis:1970yd,Zumino:1970tu,Crewther:2015dpa}).
We note the difference in sign in the $q^2$-term between the two which will play a role in  the massless dilaton case.
 We are now in a position to verify  the soft-pion theorem \eqref{eq:lthm}, 
 in each case,  
  with little effort.

\paragraph{Pions only:}  Let us suppose that the dilaton is massive and not in the low-energy theory. This corresponds 
to standard $\chi$PT for which the trace of the EMT reads (adapted to $d$-dimensions)
 \begin{equation}
 \label{eq:TEMTpi}
 \TEMT  = - \dphi (\partial \pi)^2 + \frac{d}{2} m_\pi^2 \pi^2  = - \frac{\dphi}{2} \partial^2 \pi^2 + m_\pi^2 \pi^2   \;,
  \end{equation}
  where  in the second equality the equation of motion were used 
 and it should be mentioned that only terms up to quadratic order in the pion field  were considered. 
 Taking the matrix element it is easily verified that \eqref{eq:lthm} is obeyed \cite{Donoghue:1991qv}.
 \paragraph{Pions and a massive dilaton:} This case provides us with a puzzle since the task of 
 the improvement term is to remove the $- \frac{1}{2} \partial^2 \pi^2$-term which seems vital for the low-energy theorem. 
   Hence,   one might wonder 
 how the low-energy theorem \eqref{eq:lthm} can be obeyed?  One must have a look at the new ingredient which is the dilaton 
 tadpole. Using \eqref{eq:TEMTuse} and \eqref{eq:tadpole} we get
 \begin{equation}
 \label{eq:fPiLO}
 \gone^\pi(q^2) = 1 \;, \quad {\gtwo}^\pi(q^2) = \frac{2}{d-1}  \frac{ \dphi q^2 + (\Dechi- 2 \dphi ) m_\pi^2 }{q^2-m_D^2} - 1 \;,
 \end{equation} 
 whereas the trace form factor reads
 \begin{alignat}{4}
& \gzero^\pi(q^2)   &\;=\;&  2 m_\pi^2 + \dphi q^2  - q^2  \frac{ \dphi q^2 + (\Dechi- 2 \dphi ) m_\pi^2 }{q^2-m_D^2} \nonumber \\[0.1cm]  
 & &\;=\;& 
2   m_\pi^2  + ( \dphi  + (\Dechi- 2 \dphi ) \frac{m_\pi^2}{m_D^2})   q^2 
+ \ORD(q^4) \;,
\end{alignat}
where   $p \cdot p' = m_\pi^2 - \frac{1}{2} q^2$ was used and $\dphi\equiv \frac{d-2}{2}$, as previously. 
This expression is consistent with the low-energy theorem  \eqref{eq:lthm}  if and only if 
\begin{equation}
\label{eq:Dechi}
\Dechi = d-2 \;,
\end{equation} 
holds. This is an important result  and  has been obtained in  many different ways in the context of an IRFP scenario \cite{Zwicky:2023bzk,Zwicky:2023krx}.\footnote{In the notation of those papers one has 
$\Dechi = d-1 - \gast$ and thus $\gast =1$.} The low-energy theorem can thus be regarded 
as yet another way to establish this finding. For clarity let us restate the form factors with \eqref{eq:Dechi} applied
\begin{equation}
{\gtwo}^\pi(q^2) = \frac{2}{d-1}  \frac{ \dphi q^2  }{q^2-m_D^2} - 1 \;, 
\quad \gzero^\pi(q^2)  =  2 m_\pi^2 + \dphi q^2  -   \frac{ \dphi q^4  }{q^2-m_D^2} \;.
\end{equation}
In fact the $\gzero^\pi(q^2)$ form factor has been obtained in \cite{Donoghue:1991qv} in the chiral limit assuming 
the low-energy theorem, $\sig$-meson (taking the role of the dilaton) 
dominance and convergence properties of dispersion relations. 
Hence,  chiral symmetry and assuming $\sig$-dominance is a strong enough constraint to obtain the result without 
a Lagrangian and reference to conformal symmetry. 

\paragraph{Dilaton $\gzero^D(q^2)$:}  We aim to give the dilaton form factor here and explain why 
 the soft theorem does not apply unless $\De_{T_{\mu\nu}} = d$.  Using the formulae we get at LO
 \begin{equation}
 \label{eq:fDLO}
  {\gone}^D(q^2)  = 1\;, \quad  {\gtwo}^D(q^2) =  \frac{2}{d-1}  \frac{  - \dphi q^2 +  (c_3 - 2 \dphi) m_D^2 }{q^2-m_D^2} + \frac{d-3}{d-1} \;, 
   \end{equation}
where the constant factor $\frac{d-3}{d-1}  = \frac{2(d-2)}{d-1} -1$ consists of the sum from the improvement contact term 
and the standard minus one arising from the  boson  kinetic part.  The trace form factor reads
 \begin{alignat}{4}
 \label{eq:T0D}
& \gzero^D(q^2)   &\;=\;&  2 m_D^2 - \dphi q^2  - q^2   \frac{  - \dphi q^2 +  (c_3 - 2 \dphi) m_D^2 }{q^2-m_D^2}
 \nonumber \\[0.1cm]  
 & &\;=\;& 
 2   m_D^2  + ( c_3 -  3  \dphi )   q^2 
+ \ORD(q^4) \;.
\end{alignat}
To this end we return to the question whether the   low-energy theorem \eqref{eq:lthm} could apply to massive dilatons. Not in general since the 
 commutator $[Q_D,T_{\mu\nu}]$ only vanishes when 
$\De_{T_{\mu\nu}} =d$ (excluding unforeseen cancellations). 
 This can be inferred from the soft-dilaton theorem  $F_D^2 \matel{D}{{\cal O}}{D} =  \De_{\cal O}(d-\De_{\cal O}) \vev{\cal{O}}$, 
 derived in   \cite{Zwicky:2023bzk}.  However, 
 in the  context of the potential \eqref{eq:V}, we have  $c_3 =  (d+ \De) $,
whereas the low-energy theorem requires $c_3 = 4 \dphi$ in order to match \eqref{eq:T0D}  and \eqref{eq:lthm}.
 We therefore conclude that a single operator ${\cal O}$ with  $\De_{\cal O} = d$ cannot give mass to the dilaton either. 
This is in agreement with the stronger result that if the dilaton receives a mass from a single operator ${\cal O}$ then its scaling dimension must be $\De_{\cal O} = d-2$ \cite{Zwicky:2023krx} (of which \eqref{eq:Dechi} is a special case).

 \subsection{Massless Goldstones (pion and dilaton)} 
 \label{sec:massless}
 
  Dolgov and Voloshin  \cite{Voloshin:1982eb} addressed the question 
 of whether Goldstone bosons can be improved. They required  $ \matel{\pi(p')}{\TEMT}{\pi(p)}|_{\text{LO}} =0$
which implies the \emph{conformality constraint} 
 \begin{equation}
 \label{eq:g20}
  {\gtwo}^\pi(0)= -\frac{1}{d-1} \;,
 \end{equation}
 and  additionally obtained  ${\gtwo}^\pi(0)= -1$ through the  soft-pion theorem 
  which led them to conclude that   ``Goldstone bosons
 due to internal symmetry cannot be improved". Where is the loophole in our case? It is the remainder \eqref{eq:R}.  
It  is not negligible when there is a massless dilaton and invalidates the naive use of the soft-pion theorem.
 This also explains the seemingly contradictory limit in 
 \eqref{eq:lthm}.  We may  verify the constraint \eqref{eq:g20} directly, using our results  
 \eqref{eq:fPiLO} and \eqref{eq:fDLO} 
 \begin{equation}
  {\gtwo}^\pi(0) =   \frac{2 d_\phi}{d-1} -  1 = -  \frac{1}{d-1} \;, \quad 
  {\gtwo}^D(0) = -  \frac{2 d_\phi}{d-1} +  \frac{d-3}{d-1}  = -  \frac{1}{d-1}  \;.
 \end{equation}
 We see how the different sign in the $q^2$-term in $g_{D\pi\pi}$ and $g_{DDD}$ are taken care of 
 by the contact term  see \FIG\ref{fig:dia}.

 \section{\RZ{Summary and Conclusions} }
 \label{sec:conc}
 
In this paper, we have discussed the impact of the dilaton improvement term on both Goldstone and non-Goldstone particles of spin zero and one-half. Our main results are: (1) the theoretical motivation for this improvement term in the context of spontaneous scale symmetry breaking, and (2) its effects on gravitational form factors. We calculated these form factors using a comprehensive approach that incorporates Weyl-gauging for general particles and includes the effects of soft symmetry-breaking perturbations.

The dilaton improvement term \eqref{eq:Lchi} makes the kinetic part conformal, thereby realising the hidden symmetry. While flat-space scattering amplitudes remain unaffected, the spin-zero component of the energy–momentum tensor is modified in an essential way. In particular, it realises the fundamental Goldstone matrix element \eqref{eq:FD} within the effective theory and generates the dilaton pole in ${\gtwo}(q^2)$, the gravitational form factor associated with pressure.\footnote{Identifying the classes of theories that exhibit such a phase is beyond the scope of this work. Further remarks can be found in \SEC\ref{sec:SSB}.} This pole is also closely linked to the notion of ``massive hadrons in a conformal phase,'' expressed by $\matel{\phi}{\TEMT}{\phi} = 0$ with $m_\phi \neq 0$, where the dilaton plays a role analogous to that of the pion in restoring the chiral Ward identity (Goldberger–Treiman relation) \cite{DelDebbio:2021xwu}. The transition to the standard relation $\matel{\phi}{\TEMT}{\phi}=2 m_\Phi^2$, valid for $m_D \neq 0$, has been analysed 
by regularising with momentum transfer, see \SEC\ref{sec:NGFF}.

  The pion, dilaton, scalar, and fermion form factors were obtained at LO in dilaton chiral perturbation theory,
   \begin{alignat}{4}
& {\gtwo}^\pi(q^2) &\;=\;&  \frac{2F_D}{d-1}  \frac{  g_{D\pi\pi}(q^2)   }{q^2-m_D^2} - 1 \;,  \quad 
&  & {\gtwo}^D(q^2) &\;=\;& \frac{2 F_D}{d-1}  \frac{  g_{DDD}(q^2)  }{q^2-m_D^2} + \frac{d-3}{d-1}  \;,  
\nonumber  \\[0.1cm]
\label{eq:ord}
&  {\gtwo}^\scal(q^2) &\;=\;&  \frac{2 F_D}{d-1}  \frac{  g_{D\scalg\scalg }(q^2)  }{q^2-m_D^2} - 1 \;,  \quad & & {\gtwo}^\ferm(q^2) &\;=\;& \frac{4 m_\ferm F_D}{d-1}  \frac{ \, g_{D\ferm\ferm}(q^2) }{q^2-m_D^2}  \;.
\end{alignat}
The non-Goldstone cases are described by Weyl-gauging, with $\partial_\mu D$ acting as the gauge field, though this affects only the $\ORD(q^2)$-terms in the numerator. For bosons, the standard $-1$ arises from the kinetic term, while the dilaton receives an additional contribution from the improvement term. This term controls the pole structure together with the cubic on-shell couplings,  
\begin{alignat}{4}
& g_{D\pi\pi}(q^2)   &\;=\;&   \frac{1}{F_D} (   (\Dechi -  2 \dphi) m_\pi^2+ \dphi q^2 ) \;,  \quad 
&   & g_{DDD}(q^2)  &\;=\;&  \frac{1}{F_D}  \left(     (c_3 - 2 \dphi) m_D^2-\dphi  q^2  \right) \;,
\nonumber  \\[0.1cm]
&  g_{D\scalg\scalg }(q^2)   &\;=\;& \frac{1}{F_D} (2  \bar{m}_{\scalg}^2 +  \ga_{\cal O}  \De m_{\scalg}^2  +  \dphi q^2)  \;, \quad 
& &  g_{D\ferm\ferm}(q^2) &\;=\;&  \frac{1 }{F_D} ( \bar{m}_\ferm+  \ga_{\cal O} \De m_{\ferm}) \;.
\end{alignat} 
Here $\dphi = (d-2)/2$ denotes the dimension of the free scalar, and $c_3$ is the cubic term in the dilaton potential, as discussed around \EQ\eqref{eq:LDDD}. For the scalar and fermion, a soft perturbation $\de \Lag = -\pert {\cal O}$ has been considered, with $m_\ferm = \bar{m}_\ferm + \De m_\ferm$, where the latter denotes the $\la$-induced correction, see \APP\ref{app:soft}. 
%For familiarity, the pion case was analysed with $\pert {\cal O} = m_q \bar qq$, which differs since pions, being Goldstone bosons, couple differently. 
At LO, the remaining form factors satisfy $A(q^2) =1$ and $J^\ferm(q^2) = \tfrac{1}{2}$.
In the case of a massless dilaton, the $D$-form factors satisfy the LO conformality constraints \eqref{eq:g20} and \eqref{eq:conf},
\begin{equation}
{\gtwo}^{\pi,D}(0)\big|_{m_D=0} = - \frac{1}{d-1} \;, \quad  {\gtwo}^{\scal,\ferm}(q^2)\big|_{m_D=0} = \frac{2}{d-1}  
 \frac{ 2 m_{\scal,\ferm}^2 }{q^2}   \;.
\end{equation}
The pion form factor further satisfies the stronger low-energy theorem \eqref{eq:lthm}, provided the scaling dimension of the operator $\bar qq$, which breaks chiral symmetry spontaneously, is $\De_{\bar qq} = d-2$. This result aligns with previous findings obtained through different methods \cite{Zwicky:2023bzk,Zwicky:2023krx}, under the assumption of an infrared fixed point.  
 
 The dilaton improvement mechanism explored here opens new avenues for investigating conformal dynamics through gravitational form factors, with potential applications to theories exhibiting infrared fixed points \cite{170}.

\paragraph{Acknowledgments:} 
RZ is supported by a CERN associateship and
an STFC Consolidated Grant, ST/P0000630/1.   
I am grateful to  Latham Boyle, John Ellis, Jos\'e-Ramon Espinosa, Chris Hill, George Karananas,  
Heiri Leutwyler, J\'er\'emie Quevillon, 
Christopher Smith, Lewis Tunstall, Neil Turok,  Jorinde Van de Vis, Jens-Uwe Wiese,
Sasha Zhiboedov  and the participants of the ITP Bern seminar 
 for useful correspondence and or discussions.  I am indebted to Roy Stegeman for 
 a thoroughly proofreading of these notes.  

\paragraph{Addendum:} This paper has been considerably reworked with respect to the original arXiv version with regard to gravitational 
form factors. Additionally, we have  adopted the most standard conventions in the literature: using $A$ and $\gtwo$ for the form factors and $q= p'-p$ for the momentum transfer, added the discussion 
around the Weyl-gauging and considered the effect of soft perturbations.}

\appendix

\section{Soft breaking of conformal symmetry}
\label{app:soft}

We now consider the effect of perturbing the fundamental Lagrangian by a soft operator ${\cal O}$, with scaling dimension $\Delta_{\cal O} = d_{\cal O} + \ga_{\cal O}  < d $. This leads to shifts in the mass parameters and in the gravitational form factors which we assess through the $g_{DXX}$-type couplings.  
We focus on the effect on the non-Goldstone sector, taking 
into account the  matching 
on the mass operators:
\begin{equation}
\label{eq:match}
\de \Lag =  -  \pert{\cal O} \quad \to \quad  \Lag_{\De m} =   -   
 \frac{1}{2}  \hat{\chi}^{\De_{\cal O}  - 2 \cw_{\scalg}} \,  \De m_{\scalg}^2 \, \scalg^2 
-  \hat{\chi}^{\De_{\cal O} - 2 \cw_{\ferm}}  \, \De m_{\ferm}\,  \bar{\ferm} \ferm 
 \;,
\end{equation}
where 
\begin{equation}
\De m^2_{\scalg}  = \pert  \matel{\scalg}{{\cal O}}{\scalg} \;,\quad 
\De m_{\ferm}  =  \frac{\pert}{2 m_\ferm}  \matel{\ferm}{{\cal O}}{\ferm} \;,
\end{equation}
parameterise the   mass shifts linear in $\pert$
\begin{equation}
m^2_{\scalg} = \bar{m}^2_{\scalg} +  \De m^2_{\scalg} \;, \quad 
 m_{\ferm} = \bar{m}_{\ferm} +  \De m_{\ferm}  \;.
\end{equation}
 The matrix elements 
are of zero momentum transfer such that $ \matel{\ferm}{ \bar \ferm \ferm }{\ferm} = 2 m_{\ferm}$ 
explains the extra factor in the fermion case. 

Before turning to the $g_{DXX}$-couplings, it is instructive to consider the effect on the
trace anomaly and the dilaton mass. 
 It is straightforward to verify that 
under a Weyl transformation,
\begin{equation}
 \sqrt{-g}  \Lag_{\De m} \to e^{ (d- \De_{\cal O})  \al}  \sqrt{-g}   \Lag_{\De m}  \;,
\end{equation}
which leads to a breaking of scale invariance of the form
\begin{equation}
 \TEMT  \supset  - (d- \De_{\cal O})   \Lag_{\De m} \;.
\end{equation}
We may apply this to the mass perturbation $\de \Lag = - m_q \bar qq$ in a gauge theory.
The scaling dimension is  $\De_{\bar qq}  =  (d-1)-\ga_*$ with $\ga_* = - \ga_{\bar qq}|_{\mu =0}$. 
This reproduces indeed the standard formula   $\TEMT  \supset   (1 + \ga_*) m_q \bar qq$.

%By a double-soft dilaton theorem \cite{Zwicky:2023krx}, one then finds that the dilaton mass 
%\begin{equation}
%2 \De m_D^2 = \pert(d- \De_{\cal O})  \matel{ D}{{\cal O}}{D} = - \frac{1}{F_D^2} \pert(d- \De_{\cal O})^2\De_{\cal O} \vev{\cal O} \;,
%\end{equation}
%correction to LO in the parameter $\pert$, based on the general formula 
%$2   m_D^2 = \matel{ D}{\TEMT}{D}$.

Finally, to study the corrections on the $g_{DXX}$-coupling, the effect on  
 the Lagrangian in \eqref{eq:LphiD} is needed which reads
\begin{equation}
\label{eq:LphiD2}
{\cal L}_{\scalg,\ferm,\chi} =  
\Lag_{\XX} +    \frac{1}{2}  \hat{\XX}^\beS  (  (\Wcov \scalg)^2 - 
( \hat{\chi}^{2} \bar{m}_\scalg^2  +  \hat{\XX} ^{\ga_{\cal O}}   \De m_{\scalg}^2)   \scalg^2  )
 +  \hat{\XX}^\beF \bar{\ferm}  (i \slashed{\Delta} -  ( \hat{\XX} \bar{m}_\ferm  + 
  \hat{ \XX}^{\ga_{\cal O}}  \De m_{\ferm} )  \ferm  
 \;.
\end{equation}
The scalar coupling computation in \eqref{eq:gDphiphi} are modified as 
follows\footnote{We stress that this formula does not apply to the Goldstone sector, compare 
with \eqref{eq:gDpipi},  
since the field are written in terms of coset fields with peculiar scaling dimensions.}
\begin{alignat}{2}
& g_{D \scalg\scalg}(q^2)  &\;=\;&  \frac{1}{F_D} ( [ - \beS (m_\scalg^2- \frac{1}{2}q^2) + q^2 \cw_{\scalg} ]  
+ (\beS+2) \bar{m}_{\scalg}^2 +  (\beS +  \ga_{\cal O}) \De m_{\scalg}^2 )   \nonumber \\[0.1cm]  
& &\;=\;& 
\frac{1}{F_D} (2  \bar{m}_{\scalg}^2 +  \ga_{\cal O}  \De m_{\scalg}^2  +  \dphi q^2) + \ORD(\pert^2)  \;,
\end{alignat}
and the corresponding one of the fermions in \eqref{eq:gDff} changes to
\begin{alignat}{2}
\label{eq:gDff}
& {g}_{D\ferm\ferm}  &\;=\;&\frac{1}{F_D}(0 -m_\ferm \beF + (\bar{m}_\ferm (\beF +1) 
+   (\beF + \ga_{\cal O}) \De m_{\ferm}    )    \nonumber \\[0.1cm]  
& &\;=\;&  \frac{1 }{F_D} ( \bar{m}_\ferm+  \ga_{\cal O} \De m_{\ferm})   + 
\ORD(\pert^2) \;.
\end{alignat}
It is remarkable that in both expressions, the Weyl-weights cancel out in the final result.

\section{{Relation to Perturbative Models}}
\label{app:earlier}

This appendix illustrates how the dilaton differs from a simple perturbative model.
  An example is given 
by the  complex $\la \phi^4$ model with global $U(1)$-symmetry.  It was used in   \cite{Hosotani:1986ga} to address the Dolgov-Voloshin no-go theorem. 
 The complex scalar is parameterised by
$\phi =  \phi_1 + i \phi_2$  and 
   $\de \Lag = - \xid R(  \phi_1^2 + \phi_2^2)$ takes on the role of the improvement term. 
    Spontaneous symmetry breaking is imposed by a negative mass term  and one may parameterise $\phi =  (v+\rho) e^{i \theta/v}$ 
   where $v = 6 m_\rho^2/\la$ is the vacuum expectation value. The 
  following gravitational form factor emerges 
\begin{equation}
\label{eq:HNR}
{\Gtwo}^{\rho}(q^2)|_{\text{HNR}} =  4\xid \frac{q^2}{q^2- 2 m_\rho^2} - A^\rho(q^2) \;,
\end{equation}
where  $4 G(q^2) ={\Gtwo}(q^2)+ \gone(q^2)$  translates from  \eqref{eq:RZ} to the 
basis in \cite{Hosotani:1986ga}.
They argued that  the  soft theorem
constraint  \eqref{eq:softg}  is obeyed 
 and improvement  was taking place in terms of $\de \Lag$   \cite{Hosotani:1986ga}.   
Note that the conformality constraint cannot be fulfilled since $m_\rho \neq 0$. 
 How does this relate to this  work?
\begin{itemize}
\item [-] If $m_\rho, \la \to 0$ for  $v \neq 0$, then the radial mode  
$\rho$ and the angle $\theta$  take on the roles of the dilaton  and the 
``pion" due to the internal symmetry breaking.  In this setting the solutions are equivalent. 
However, beyond LO this correspondence would require 
fine tuning.
\item [-] One may wonder what happens when the global symmetry is enlarged, 
 such as  in the linear $\sig$-model: $U(N_f) \otimes U(N_f) \to U(N_f)$.
For $N_f > 2$ the analogy  breaks down as there are generally $2 N_f^2$ degrees of freedom of 
which $N_f^2$ become massless (see \cite{Paterson:1980fc}).\footnote{For $N_f =3$ the eighteen particles  correspond to the set of  nine 
parity-even  $\{ \sig, 3 \times a_0(980) ,4\times K^*_0(700),  f_0(980)\}$-  and nine  parity-odd 
 $\{ \eta', 3 \times \pi, 4 \times K ,\eta\}$-mesons.}
 Hence tuning the mass to zero in those models would lead 
to $N_f^2$ apparent dilatons. 
An exception is the much studied $N_f=2$ case,  the original $\sig$-model, for which 
the representation is reducible  due to the pseudo-reality of $SU(2)$ \cite{Paterson:1980fc}. 
The spectrum then consists of one $\sig$-meson and three pions. 
 The $\sig$-meson can take on the role of the dilaton in the fine tuned case.
 \end{itemize}
 In summary, in certain perturbative models the $\sig$-particle (or Higgs) can take on the role 
  of the dilaton when fine-tuned.

\section{Renormalisation Group Matters}
\label{app:RG}

\subsection{On the running $\xi(\mu)$}
 \label{app:run}

 Let us discuss  renormalisation group effects on $\xi(\mu)$ and then turn to the particular role 
 at FPs. 
 The non-minimal coupling $\xi$ in \eqref{eq:Lag} cannot be ignored since 
 it will generally appear through renormalisation  effects, see  
 \cite{Freedman:1974ze,Collins:1976va,Brown:1980qq,Muta:1991mw}. The expression in \eqref{eq:dphi} should be seen as a LO-value, corrected by 
 an expansion in a coupling constant $\la$, order by order in perturbation theory, 
 $\xi(\mu) = \xid + \De \xi(\la(\mu))$. 
 At trivial FPs $ \De \xi \to 0$ which is automatic since 
 the coupling  approaches zero. This leads to all the good properties such as  
 UV-finiteness and compatibility with the weak equivalence principle in the deep-IR.   
 
Concretely, the renormalisation and the renormalisation group  equation were first studied in  $\la \phi^4$-models  \cite{Freedman:1974ze,Collins:1976va,Brown:1980qq} where it was found that 
non-finite counterterms enter at $\ORD(\la^3)$.  In fact the absence of such terms would 
imply the existence of an unknown or hidden symmetry \cite{Freedman:1974ze}, which would have 
been an  exceptional circumstance.
The counterterms render the renormalisation group equation non-homogeneous. This means that 
$\De \xi(\mu) =0$ cannot be consistently imposed and that $\xi = \xi_d$ is not a renormalisation group FP, 
in the interacting theory. 
In the $\la \phi^4$-model,  $ \De \xi \to 0$ at the IRFP is discussed in \cite{Brown:1980qq}. 
For the UV, the situation is unclear since $\la \phi^4$ has either the triviality problem \cite{Fernandez:1992jh} or
an unknown non-perturbative FP.
The situation in this paper concerns Goldstones which are specific fields of an effective theory which is IR-free. One can therefore expect that for   $\De \xi(\mu) \to 0$ for $\mu \to 0$ to holds.

 In summary, $\xi(\mu)$  ought to assume its free field value at trivial FPs.
Outside FPs it is RG-scale dependent and as such 
renormalisation scheme-dependent quantities.
 
\subsection{\RZ{Flow Theorems and  Improvement Terms}}
\label{app:flow}

Flow theorems state a hierarchy between the UV and the IR limit of the Weyl anomaly coefficients for theories
flowing from a UV to an IRFP (or CFT). Originally formulated in 
 $d=2$, known as the $c$-theorem,  by Zamalodchikov \cite{Zamolodchikov:1986gt},  Cardy \cite{Cardy:1988tj} provided an insightful  formula 
\begin{equation}
\label{eq:cthm}
c_{\text{UV}} - c_{\text{IR}}  = 3 \pi  \int d^2 x x^2 \vev{\TEMT(x)\TEMT(0)} > 0 \;,
\end{equation}
where $c$ is the central charge  in the Weyl anomaly $\vev{\TEMT} = - \frac{1}{24 \pi} c R$.  One can infer from \eqref{eq:cthm} that  for the integral to converge it is 
 important that  $\TEMT \to 0$ fast enough in the UV \emph{and} the IR.

In $d=4$ the Weyl anomaly reads $\vev{\TEMT} =   -( \be_a^* E_4 + \be_b^* R^2  +  \be_c^* W^2  ) + \frac{4}{3} \bar b^* \Box R$ (closely following the conventions in \cite{Prochazka:2017pfa}) with the asterisk denoting the IR limit such as 
$\be_a^* = \be_a|_{\mu=0}$. The well known   $a$-theorem concerns the flow of 
the Euler term $E_4$ \cite{Cardy:1988cwa,Jack:1990eb,Komargodski:2011vj,Luty:2012ww,Shore:2016xor} but 
the $\Box R$-term is also a candidate \cite{Prochazka:2017pfa}.\footnote{With regard to the latter the main point is that
whereas the $\Box R$-term can be changed by adding a local counterterm in a specific theory it would cancel in a difference such as in \eqref{eq:cthm}.}  
UV-convergence can be shown to hold  by renormalisation group   resummation  techniques  \cite{Prochazka:2016ati}.
Below we explain why non-improved pions are problematic in the IR. 

Let us start with the $\Box R$-term for which the flow theorem takes on a very analogous formula to \eqref{eq:cthm}
\begin{equation}
\label{eq:bthm}
\bar b_{\text{UV}} - \bar b_{\text{IR}}  =  \frac{1}{3 \, 2^9}  \int d^4 x\,  x^4 \vev{\TEMT(x)\TEMT(0)} > 0 \;,
\end{equation}
in that it is expressible in terms of a two-point function.   If the IR theory is standard QCD with free pions in the IR, then 
the non-improved pion theory  
  $\TEMT = - \frac{1}{2} \partial^2 \pi^2$ leads to  $ \vev{\TEMT(x)\TEMT(0)} \propto{1}/{x^8}$ which 
  is logarthimcally divergent in the IR  \cite{Prochazka:2017pfa}. In that reference the discussion is phrased in 
  terms of the momentum space or dispersion representation but the outcome is the same.
  
Let us turn to the $a$-theorem, the flow of $E_4$,   for which the   two-point function is no longer sufficient.   
However, the same two-point function is still relevant  since it contributes to the flow. 
 We  use the language in \cite{Komargodski:2011vj,Luty:2012ww}.
 A pathway to positivity is given by unitarity and the four-point function scattering amplitude ${\cal A}(s)$ of 
 four \emph{external}  dilaton fields $\tau$ which are put ``'on-shell" $\partial^2 \tau =0$.  The difference of the Weyl anomaly coefficients 
  $\be_a^{\text{UV}} - \be_a^{\text{IR}} \propto \al(\infty) - \al(0)$ is directly related to the scattering amplitude  by 
   ${\cal A}(s) = s^2 \al(s)/f^4$, where $f$  the respective dilaton decay constant. 
 One finds (see \EQ 3.7 in \cite{Luty:2012ww})
  \begin{equation}
  f^4 {\cal A}(s)  = \vev{\TEMT(p_1+p_2) \TEMT(p_3+p_4)} + \dots
  \end{equation}
  with $p_i$ denoting the dilaton momenta, 
   $s \equiv (p_1+p_2)^2$ the centre of mass energy and the dots stand for other contributions, including  higher point-functions. 
With   $\TEMT = - \frac{1}{2} \partial^2 \pi^2$ this leads to $ {\cal A}(s) \supset s^2 \ln s$ and  $\Ima \al(s) = c_0 + \dots$.
  Since $\be_a^{\text{UV}} - \be_a^{\text{IR}} \supset  c' \int_0^\infty \Ima \al(s')/s' ds'$ the same type of logarithmic divergence is found as for the $\Box R$-flow discussed above. Note that for all other correlation functions the on-shell condition $p_i^2 =0$ 
  removes the  $ - \frac{1}{2} \partial^2 \pi^2$-term. In essence it is that $p_i^2=0$ does not imply $(p_1+p_2)^2=0$ 
  which leads to the apparent non-convergence. 

In summary if the system can be improved, with a \emph{dynamical} dilaton, 
then these problems dissolve.\footnote{If a massless dilaton were present then the Goldstone counting, as first applied 
in \cite{Cardy:1988cwa}, would be modified by the addition of the dilaton.  In practice the  bounds would not change significantly 
since the there are many other particles entering the inequality. }  If that was not the case then this does still not mean that 
the flow-theorems do not hold. 
It could just be that the formulae to compute them
break down in the Goldstone phase and need  amendment.  For example it could be 
that one needs to focus on the purely anomalous part and regard  $\TEMT = - \frac{1}{2} \partial^2 \pi^2$
as explicit symmetry breaking as briefly discussed in \cite{Larue:2023qxw}.

\bibliographystyle{utphys}
\bibliography{../Dil-refs.bib}

\providecommand{\href}[2]{#2}\begingroup\raggedright\begin{thebibliography}{10}

\bibitem{Callan:1970ze}
C.~G. Callan, Jr., S.~R. Coleman, and R.~Jackiw, ``{A New improved energy -
  momentum tensor},''
  \href{http://dx.doi.org/10.1016/0003-4916(70)90394-5}{{\em Annals Phys.}
  {\bfseries 59} (1970) 42--73}.

\bibitem{DeWitt:1975ys}
B.~S. DeWitt, ``{Quantum Field Theory in Curved Space-Time},''
  \href{http://dx.doi.org/10.1016/0370-1573(75)90051-4}{{\em Phys. Rept.}
  {\bfseries 19} (1975) 295--357}.

\bibitem{Chernikov:1968zm}
N.~A. Chernikov and E.~A. Tagirov, ``{Quantum theory of scalar fields in de
  Sitter space-time},'' {\em Ann. Inst. H. Poincare Phys. Theor. A} {\bfseries
  9} (1968) 109.

\bibitem{Voloshin:1982eb}
M.~B. Voloshin and A.~D. Dolgov, ``{ON GRAVITATIONAL INTERACTION OF THE
  GOLDSTONE BOSONS},'' {\em Sov. J. Nucl. Phys.} {\bfseries 35} (1982)
  120--121.

\bibitem{Leutwyler:1989tn}
H.~Leutwyler and M.~A. Shifman, ``{GOLDSTONE BOSONS GENERATE PECULIAR CONFORMAL
  ANOMALIES},'' \href{http://dx.doi.org/10.1016/0370-2693(89)91730-9}{{\em
  Phys. Lett. B} {\bfseries 221} (1989) 384--388}.

\bibitem{Donoghue:1990xh}
J.~F. Donoghue, J.~Gasser, and H.~Leutwyler, ``{The Decay of a Light Higgs
  Boson},'' \href{http://dx.doi.org/10.1016/0550-3213(90)90474-R}{{\em Nucl.
  Phys. B} {\bfseries 343} (1990) 341--368}.

\bibitem{Donoghue:1991qv}
J.~F. Donoghue and H.~Leutwyler, ``{Energy and momentum in chiral theories},''
  \href{http://dx.doi.org/10.1007/BF01560453}{{\em Z. Phys. C} {\bfseries 52}
  (1991) 343--351}.

\bibitem{Isham:1970gz}
C.~J. Isham, A.~Salam, and J.~A. Strathdee, ``{Spontaneous breakdown of
  conformal symmetry},''
  \href{http://dx.doi.org/10.1016/0370-2693(70)90177-2}{{\em Phys. Lett. B}
  {\bfseries 31} (1970) 300--302}.

\bibitem{Coleman:1985rnk}
S.~Coleman, \href{http://dx.doi.org/10.1017/CBO9780511565045}{{\em {Aspects of
  Symmetry}: {Selected Erice Lectures}}}.
\newblock Cambridge University Press, Cambridge, U.K., 1985.

\bibitem{Low:2001bw}
I.~Low and A.~V. Manohar, ``{Spontaneously broken space-time symmetries and
  Goldstone's theorem},''
  \href{http://dx.doi.org/10.1103/PhysRevLett.88.101602}{{\em Phys. Rev. Lett.}
  {\bfseries 88} (2002) 101602},
  \href{http://arxiv.org/abs/hep-th/0110285}{{\ttfamily arXiv:hep-th/0110285}}.

\bibitem{Migdal:1982jp}
A.~A. Migdal and M.~A. Shifman, ``{Dilaton Effective Lagrangian in
  Gluodynamics},'' \href{http://dx.doi.org/10.1016/0370-2693(82)90089-2}{{\em
  Phys. Lett. B} {\bfseries 114} (1982) 445--449}.

\bibitem{Schwimmer:2010za}
A.~Schwimmer and S.~Theisen, ``{Spontaneous Breaking of Conformal Invariance
  and Trace Anomaly Matching},''
  \href{http://dx.doi.org/10.1016/j.nuclphysb.2011.02.003}{{\em Nucl. Phys. B}
  {\bfseries 847} (2011) 590--611},
  \href{http://arxiv.org/abs/1011.0696}{{\ttfamily arXiv:1011.0696 [hep-th]}}.

\bibitem{Komargodski:2011vj}
Z.~Komargodski and A.~Schwimmer, ``{On Renormalization Group Flows in Four
  Dimensions},'' \href{http://dx.doi.org/10.1007/JHEP12(2011)099}{{\em JHEP}
  {\bfseries 12} (2011) 099}, \href{http://arxiv.org/abs/1107.3987}{{\ttfamily
  arXiv:1107.3987 [hep-th]}}.

\bibitem{Bobev:2013vta}
N.~Bobev, H.~Elvang, and T.~M. Olson, ``{Dilaton effective action with N = 1
  supersymmetry},'' \href{http://dx.doi.org/10.1007/JHEP04(2014)157}{{\em JHEP}
  {\bfseries 04} (2014) 157}, \href{http://arxiv.org/abs/1312.2925}{{\ttfamily
  arXiv:1312.2925 [hep-th]}}.

\bibitem{Golterman:2016lsd}
M.~Golterman and Y.~Shamir, ``{Low-energy effective action for pions and a
  dilatonic meson},'' \href{http://dx.doi.org/10.1103/PhysRevD.94.054502}{{\em
  Phys. Rev. D} {\bfseries 94} no.~5, (2016) 054502},
  \href{http://arxiv.org/abs/1603.04575}{{\ttfamily arXiv:1603.04575
  [hep-ph]}}.

\bibitem{Zwicky:2023krx}
R.~Zwicky, ``{QCD with an infrared fixed point and a dilaton},''
  \href{http://dx.doi.org/10.1103/PhysRevD.110.014048}{{\em Phys. Rev. D}
  {\bfseries 110} no.~1, (2024) 014048},
  \href{http://arxiv.org/abs/2312.13761}{{\ttfamily arXiv:2312.13761
  [hep-ph]}}.

\bibitem{Yamawaki:2015tmu}
K.~Yamawaki, ``{Old wine in a new bottle: Technidilaton as the 125 GeV
  Higgs},'' \href{http://dx.doi.org/10.1142/S0217751X17470261}{{\em Int. J.
  Mod. Phys. A} {\bfseries 32} no.~36, (2017) 1747026},
  \href{http://arxiv.org/abs/1511.06883}{{\ttfamily arXiv:1511.06883
  [hep-ph]}}.

\bibitem{Appelquist:2022mjb}
T.~Appelquist, J.~Ingoldby, and M.~Piai, ``{Dilaton Effective Field Theory},''
  \href{http://dx.doi.org/10.3390/universe9010010}{{\em Universe} {\bfseries 9}
  no.~1, (2023) 10}, \href{http://arxiv.org/abs/2209.14867}{{\ttfamily
  arXiv:2209.14867 [hep-ph]}}.

\bibitem{Zwicky:2023bzk}
R.~Zwicky, ``{QCD with an infrared fixed point: The pion sector},''
  \href{http://dx.doi.org/10.1103/PhysRevD.109.034009}{{\em Phys. Rev. D}
  {\bfseries 109} no.~3, (2024) 034009},
  \href{http://arxiv.org/abs/2306.06752}{{\ttfamily arXiv:2306.06752
  [hep-ph]}}.

\bibitem{Feynman:1996kb}
R.~P. Feynman, {\em {Feynman lectures on gravitation}}.
\newblock 1996.

\bibitem{Misner:1973prb}
C.~W. Misner, K.~S. Thorne, and J.~A. Wheeler, {\em {Gravitation}}.
\newblock W. H. Freeman, San Francisco, 1973.

\bibitem{DiFrancesco:1997nk}
P.~Di~Francesco, P.~Mathieu, and D.~Senechal,
  \href{http://dx.doi.org/10.1007/978-1-4612-2256-9}{{\em {Conformal Field
  Theory}}}.
\newblock Graduate Texts in Contemporary Physics. Springer-Verlag, New York,
  1997.

\bibitem{DeWitt:1964oba}
R.~Penrose, ``{Relativit\'e, Groupes et Topologie}: {Proceedings, Ecole
  d'\'et\'e de Physique Th\'eorique, Session XIII, Les Houches, France, Jul 1 -
  Aug 24, 1963},''
\newblock Gordon and Breach, New York, 1964.

\bibitem{Grib:1995xm}
A.~A. Grib and E.~A. Poberii, ``{On the difference between conformal and
  minimal couplings in general relativity},'' {\em Helv. Phys. Acta} {\bfseries
  68} (1995) 380--395.

\bibitem{Donoghue:1992dd}
J.~F. Donoghue, E.~Golowich, and B.~R. Holstein,
  \href{http://dx.doi.org/10.1017/CBO9780511524370}{{\em {Dynamics of the
  standard model}}}, vol.~2.
\newblock CUP, 2014.

\bibitem{Scherer:2012xha}
S.~Scherer and M.~R. Schindler,
  \href{http://dx.doi.org/10.1007/978-3-642-19254-8}{{\em {A Primer for Chiral
  Perturbation Theory}}}, vol.~830.
\newblock 2012.

\bibitem{Goldstone:1962es}
J.~Goldstone, A.~Salam, and S.~Weinberg, ``{Broken Symmetries},''
  \href{http://dx.doi.org/10.1103/PhysRev.127.965}{{\em Phys. Rev.} {\bfseries
  127} (1962) 965--970}.

\bibitem{Karananas:2015ioa}
G.~K. Karananas and A.~Monin, ``{Weyl vs. Conformal},''
  \href{http://dx.doi.org/10.1016/j.physletb.2016.04.001}{{\em Phys. Lett. B}
  {\bfseries 757} (2016) 257--260},
  \href{http://arxiv.org/abs/1510.08042}{{\ttfamily arXiv:1510.08042
  [hep-th]}}.

\bibitem{Nakayama:2013is}
Y.~Nakayama, ``{Scale invariance vs conformal invariance},''
  \href{http://dx.doi.org/10.1016/j.physrep.2014.12.003}{{\em Phys. Rept.}
  {\bfseries 569} (2015) 1--93},
  \href{http://arxiv.org/abs/1302.0884}{{\ttfamily arXiv:1302.0884 [hep-th]}}.

\bibitem{Fortin:2012hn}
J.-F. Fortin, B.~Grinstein, and A.~Stergiou, ``{Limit Cycles and Conformal
  Invariance},'' \href{http://dx.doi.org/10.1007/JHEP01(2013)184}{{\em JHEP}
  {\bfseries 01} (2013) 184}, \href{http://arxiv.org/abs/1208.3674}{{\ttfamily
  arXiv:1208.3674 [hep-th]}}.

\bibitem{Luty:2012ww}
M.~A. Luty, J.~Polchinski, and R.~Rattazzi, ``{The $a$-theorem and the
  Asymptotics of 4D Quantum Field Theory},''
  \href{http://dx.doi.org/10.1007/JHEP01(2013)152}{{\em JHEP} {\bfseries 01}
  (2013) 152}, \href{http://arxiv.org/abs/1204.5221}{{\ttfamily arXiv:1204.5221
  [hep-th]}}.

\bibitem{Bzowski:2014qja}
A.~Bzowski and K.~Skenderis, ``{Comments on scale and conformal invariance},''
  \href{http://dx.doi.org/10.1007/JHEP08(2014)027}{{\em JHEP} {\bfseries 08}
  (2014) 027}, \href{http://arxiv.org/abs/1402.3208}{{\ttfamily arXiv:1402.3208
  [hep-th]}}.

\bibitem{Dymarsky:2014zja}
A.~Dymarsky, K.~Farnsworth, Z.~Komargodski, M.~A. Luty, and V.~Prilepina,
  ``{Scale Invariance, Conformality, and Generalized Free Fields},''
  \href{http://dx.doi.org/10.1007/JHEP02(2016)099}{{\em JHEP} {\bfseries 02}
  (2016) 099}, \href{http://arxiv.org/abs/1402.6322}{{\ttfamily arXiv:1402.6322
  [hep-th]}}.

\bibitem{Dymarsky:2015jia}
A.~Dymarsky and A.~Zhiboedov, ``{Scale-invariant breaking of conformal
  symmetry},'' \href{http://dx.doi.org/10.1088/1751-8113/48/41/41FT01}{{\em J.
  Phys. A} {\bfseries 48} no.~41, (2015) 41FT01},
  \href{http://arxiv.org/abs/1505.01152}{{\ttfamily arXiv:1505.01152
  [hep-th]}}.

\bibitem{Semenoff:2018yrt}
G.~W. Semenoff, ``{Dilaton in a cold Fermi gas},'' in {\em {7th International
  Conference on New Frontiers in Physics}}.
\newblock 8, 2018.
\newblock \href{http://arxiv.org/abs/1808.03861}{{\ttfamily arXiv:1808.03861
  [cond-mat.quant-gas]}}.

\bibitem{Bardeen:1983rv}
W.~A. Bardeen, M.~Moshe, and M.~Bander, ``{Spontaneous Breaking of Scale
  Invariance and the Ultraviolet Fixed Point in O($N$) Symmetric
  $(\bar{\phi}^6_3$ in Three-Dimensions) Theory},''
  \href{http://dx.doi.org/10.1103/PhysRevLett.52.1188}{{\em Phys. Rev. Lett.}
  {\bfseries 52} (1984) 1188}.

\bibitem{Cresswell-Hogg:2023hdg}
C.~Cresswell-Hogg and D.~F. Litim, ``{Scale Symmetry Breaking and Generation of
  Mass at Quantum Critical Points},''
  \href{http://arxiv.org/abs/2311.16246}{{\ttfamily arXiv:2311.16246
  [hep-th]}}.

\bibitem{Semenoff:2024prf}
G.~W. Semenoff and R.~A. Stewart, ``{Dilaton in a multicritical 3+epsilon-D
  parity violating field theory},''
  \href{http://dx.doi.org/10.1016/j.physletb.2024.138691}{{\em Phys. Lett. B}
  {\bfseries 853} (2024) 138691},
  \href{http://arxiv.org/abs/2402.09646}{{\ttfamily arXiv:2402.09646
  [hep-th]}}.

\bibitem{prep}
C.~Cresswell-Hogg, D.~F. Litim, , and R.~Zwicky, {\em in preparation}.

\bibitem{DelDebbio:2021xwu}
L.~Del~Debbio and R.~Zwicky, ``{Dilaton and massive hadrons in a conformal
  phase},'' \href{http://dx.doi.org/10.1007/JHEP08(2022)007}{{\em JHEP}
  {\bfseries 08} (2022) 007}, \href{http://arxiv.org/abs/2112.11363}{{\ttfamily
  arXiv:2112.11363 [hep-ph]}}.

\bibitem{Crewther:2012wd}
R.~J. Crewther and L.~C. Tunstall, ``{Origin of $\Delta I=1/2$ Rule for Kaon
  Decays: QCD Infrared Fixed Point},''
  \href{http://arxiv.org/abs/1203.1321}{{\ttfamily arXiv:1203.1321 [hep-ph]}}.

\bibitem{Crewther:2015dpa}
R.~J. Crewther and L.~C. Tunstall, ``{Status of Chiral-Scale Perturbation
  Theory},'' \href{http://dx.doi.org/10.22323/1.253.0132}{{\em PoS} {\bfseries
  CD15} (2015) 132}, \href{http://arxiv.org/abs/1510.01322}{{\ttfamily
  arXiv:1510.01322 [hep-ph]}}.

\bibitem{Ellis:1970yd}
J.~R. Ellis, ``{Aspects of conformal symmetry and chirality},''
  \href{http://dx.doi.org/10.1016/0550-3213(70)90422-0}{{\em Nucl. Phys. B}
  {\bfseries 22} (1970) 478--492}. [Erratum: Nucl.Phys.B 25, 639--639 (1971)].

\bibitem{Iorio:1996ad}
A.~Iorio, L.~O'Raifeartaigh, I.~Sachs, and C.~Wiesendanger, ``{Weyl gauging and
  conformal invariance},''
  \href{http://dx.doi.org/10.1016/S0550-3213(97)00190-9}{{\em Nucl. Phys. B}
  {\bfseries 495} (1997) 433--450},
  \href{http://arxiv.org/abs/hep-th/9607110}{{\ttfamily arXiv:hep-th/9607110}}.

\bibitem{Zumino:1970tu}
B.~Zumino, {\em {Effective Lagrangians and Broken Symmetries}}, vol.~2 of {\em
  1970 Brandeis University Summer Institute in Theoretical Physics, Vol. 2}.
\newblock (M.I.T. Press, Cambridge, MA, 1970), Providence, RI, 1970.

\bibitem{Polyakov:2018zvc}
M.~V. Polyakov and P.~Schweitzer, ``{Forces inside hadrons: pressure, surface
  tension, mechanical radius, and all that},''
  \href{http://dx.doi.org/10.1142/S0217751X18300259}{{\em Int. J. Mod. Phys. A}
  {\bfseries 33} no.~26, (2018) 1830025},
  \href{http://arxiv.org/abs/1805.06596}{{\ttfamily arXiv:1805.06596
  [hep-ph]}}.

\bibitem{Burkert:2018bqq}
V.~D. Burkert, L.~Elouadrhiri, and F.~X. Girod, ``{The pressure distribution
  inside the proton},'' \href{http://dx.doi.org/10.1038/s41586-018-0060-z}{{\em
  Nature} {\bfseries 557} no.~7705, (2018) 396--399}.

\bibitem{Duran:2022xag}
B.~Duran {\em et~al.}, ``{Determining the gluonic gravitational form factors of
  the proton},'' \href{http://dx.doi.org/10.1038/s41586-023-05730-4}{{\em
  Nature} {\bfseries 615} no.~7954, (2023) 813--816},
  \href{http://arxiv.org/abs/2207.05212}{{\ttfamily arXiv:2207.05212
  [nucl-ex]}}.

\bibitem{Lorce:2017xzd}
C.~Lorc\'e, ``{On the hadron mass decomposition},''
  \href{http://dx.doi.org/10.1140/epjc/s10052-018-5561-2}{{\em Eur. Phys. J. C}
  {\bfseries 78} no.~2, (2018) 120},
  \href{http://arxiv.org/abs/1706.05853}{{\ttfamily arXiv:1706.05853
  [hep-ph]}}.

\bibitem{Liu:2021gco}
K.-F. Liu, ``{Proton mass decomposition and hadron cosmological constant},''
  \href{http://dx.doi.org/10.1103/PhysRevD.104.076010}{{\em Phys. Rev. D}
  {\bfseries 104} no.~7, (2021) 076010},
  \href{http://arxiv.org/abs/2103.15768}{{\ttfamily arXiv:2103.15768
  [hep-ph]}}.

\bibitem{Ji:2021qgo}
X.~Ji, Y.~Liu, and A.~Sch\"afer, ``{Scale symmetry breaking, quantum anomalous
  energy and proton mass decomposition},''
  \href{http://dx.doi.org/10.1016/j.nuclphysb.2021.115537}{{\em Nucl. Phys. B}
  {\bfseries 971} (2021) 115537},
  \href{http://arxiv.org/abs/2105.03974}{{\ttfamily arXiv:2105.03974
  [hep-ph]}}.

\bibitem{Ji:2021mtz}
X.~Ji, ``{Proton mass decomposition: naturalness and interpretations},''
  \href{http://dx.doi.org/10.1007/s11467-021-1065-x}{{\em Front. Phys.
  (Beijing)} {\bfseries 16} no.~6, (2021) 64601},
  \href{http://arxiv.org/abs/2102.07830}{{\ttfamily arXiv:2102.07830
  [hep-ph]}}.

\bibitem{Hudson:2016gnq}
J.~Hudson, I.~A. Perevalova, M.~V. Polyakov, and P.~Schweitzer, ``{Structure of
  the Energy-Momentum Tensor and Applications},''
  \href{http://dx.doi.org/10.22323/1.284.0007}{{\em PoS} {\bfseries QCDEV2016}
  (2017) 007}, \href{http://arxiv.org/abs/1612.06721}{{\ttfamily
  arXiv:1612.06721 [hep-ph]}}.

\bibitem{Novikov:1980fa}
V.~A. Novikov and M.~A. Shifman, ``{Comment on the psi-prime ---\ensuremath{>}
  J/psi pi pi Decay},'' \href{http://dx.doi.org/10.1007/BF01429829}{{\em Z.
  Phys. C} {\bfseries 8} (1981) 43}.

\bibitem{Moussallam:2011zg}
B.~Moussallam, ``{Couplings of light I=0 scalar mesons to simple operators in
  the complex plane},''
  \href{http://dx.doi.org/10.1140/epjc/s10052-011-1814-z}{{\em Eur. Phys. J. C}
  {\bfseries 71} (2011) 1814}, \href{http://arxiv.org/abs/1110.6074}{{\ttfamily
  arXiv:1110.6074 [hep-ph]}}.

\bibitem{Hoferichter:2023mgy}
M.~Hoferichter, J.~R. de~Elvira, B.~Kubis, and U.-G. Mei\ss{}ner, ``{Nucleon
  resonance parameters from Roy-Steiner equations},''
  \href{http://arxiv.org/abs/2312.15015}{{\ttfamily arXiv:2312.15015
  [hep-ph]}}.

\bibitem{170}
R.~Stegeman and R.~Zwicky, {\em Gravitational $ D$-Form Factor: The
  $\sigma$-Meson as a Dilaton confronted with Lattice Data}.

\bibitem{Hosotani:1986ga}
Y.~Hosotani, M.~Nikolic, and S.~Rudaz, ``{PSEUDOGOLDSTONE BOSONS IN CURVED
  SPACE-TIME},'' \href{http://dx.doi.org/10.1103/PhysRevD.34.627}{{\em Phys.
  Rev. D} {\bfseries 34} (1986) 627}.

\bibitem{Paterson:1980fc}
A.~J. Paterson, ``{{Coleman-Weinberg} Symmetry Breaking in the Chiral SU($N$) X
  SU($N$) Linear Sigma Model},''
  \href{http://dx.doi.org/10.1016/0550-3213(81)90489-2}{{\em Nucl. Phys. B}
  {\bfseries 190} (1981) 188--204}.

\bibitem{Freedman:1974ze}
D.~Z. Freedman and E.~J. Weinberg, ``{The Energy-Momentum Tensor in Scalar and
  Gauge Field Theories},''
  \href{http://dx.doi.org/10.1016/0003-4916(74)90040-2}{{\em Annals Phys.}
  {\bfseries 87} (1974) 354}.

\bibitem{Collins:1976va}
J.~C. Collins, ``{A Finite Improvement Renormalizes the Energy-Momentum
  Tensor},'' \href{http://dx.doi.org/10.1103/PhysRevLett.36.1518}{{\em Phys.
  Rev. Lett.} {\bfseries 36} (1976) 1518}.

\bibitem{Brown:1980qq}
L.~S. Brown and J.~C. Collins, ``{Dimensional Renormalization of Scalar Field
  Theory in Curved Space-time},''
  \href{http://dx.doi.org/10.1016/0003-4916(80)90232-8}{{\em Annals Phys.}
  {\bfseries 130} (1980) 215}.

\bibitem{Muta:1991mw}
T.~Muta and S.~D. Odintsov, ``{Model dependence of the nonminimal scalar
  graviton effective coupling constant in curved space-time},''
  \href{http://dx.doi.org/10.1142/S0217732391004206}{{\em Mod. Phys. Lett. A}
  {\bfseries 6} (1991) 3641--3646}.

\bibitem{Fernandez:1992jh}
R.~Fernandez, J.~Frohlich, and A.~D. Sokal, {\em {Random walks, critical
  phenomena, and triviality in quantum field theory}}.
\newblock 1992.

\bibitem{Zamolodchikov:1986gt}
A.~B. Zamolodchikov, ``{Irreversibility of the Flux of the Renormalization
  Group in a 2D Field Theory},'' {\em JETP Lett.} {\bfseries 43} (1986)
  730--732.

\bibitem{Cardy:1988tj}
J.~L. Cardy, ``{The Central Charge and Universal Combinations of Amplitudes in
  Two-dimensional Theories Away From Criticality},''
  \href{http://dx.doi.org/10.1103/PhysRevLett.60.2709}{{\em Phys. Rev. Lett.}
  {\bfseries 60} (1988) 2709}.

\bibitem{Prochazka:2017pfa}
V.~Prochazka and R.~Zwicky, ``{On the Flow of $\Box R$ Weyl-Anomaly},''
  \href{http://dx.doi.org/10.1103/PhysRevD.96.045011}{{\em Phys. Rev. D}
  {\bfseries 96} no.~4, (2017) 045011},
  \href{http://arxiv.org/abs/1703.01239}{{\ttfamily arXiv:1703.01239
  [hep-th]}}.

\bibitem{Cardy:1988cwa}
J.~L. Cardy, ``{Is There a c Theorem in Four-Dimensions?},''
  \href{http://dx.doi.org/10.1016/0370-2693(88)90054-8}{{\em Phys. Lett. B}
  {\bfseries 215} (1988) 749--752}.

\bibitem{Jack:1990eb}
I.~Jack and H.~Osborn, ``{Analogs for the $c$ Theorem for Four-dimensional
  Renormalizable Field Theories},''
  \href{http://dx.doi.org/10.1016/0550-3213(90)90584-Z}{{\em Nucl. Phys. B}
  {\bfseries 343} (1990) 647--688}.

\bibitem{Shore:2016xor}
G.~M. Shore, \href{http://dx.doi.org/10.1007/978-3-319-54000-9}{{\em {The c and
  a-theorems and the Local Renormalisation Group}}}.
\newblock SpringerBriefs in Physics. Springer, Cham, 2017.
\newblock \href{http://arxiv.org/abs/1601.06662}{{\ttfamily arXiv:1601.06662
  [hep-th]}}.

\bibitem{Prochazka:2016ati}
V.~Prochazka and R.~Zwicky, ``{Finiteness of two- and three-point functions and
  the renormalization group},''
  \href{http://dx.doi.org/10.1103/PhysRevD.95.065027}{{\em Phys. Rev. D}
  {\bfseries 95} no.~6, (2017) 065027},
  \href{http://arxiv.org/abs/1611.01367}{{\ttfamily arXiv:1611.01367
  [hep-th]}}.

\bibitem{Larue:2023qxw}
R.~Larue, J.~Quevillon, and R.~Zwicky, ``{Gravity-gauge anomaly constraints on
  the energy-momentum tensor},''
  \href{http://dx.doi.org/10.1007/JHEP05(2024)307}{{\em JHEP} {\bfseries 05}
  (2024) 307}, \href{http://arxiv.org/abs/2312.13222}{{\ttfamily
  arXiv:2312.13222 [hep-th]}}.

\end{thebibliography}\endgroup

\end{document}